\documentclass[fleqn,usenatbib]{mnras}
\usepackage{newtxtext,newtxmath}

\usepackage[T1]{fontenc}
\usepackage{ae,aecompl}

\usepackage{graphicx}	
\usepackage{amsmath}	
\usepackage{amssymb}	

\newcommand{\MJ}{M_{\rm J}}
\newcommand{\RJ}{R_{\rm J}} 
\newcommand{\Rp}{R_{\rm p}} 
\newcommand{\rp}{r_{\rm p}} 
\newcommand{\Mk}{M^{(k)}}
\newcommand{\xCMk}{{\bf x}_{\rm CM}^{(k)}} 
\newcommand{\vCMk}{{\bf v}_{\rm CM}^{(k)}} 

\title[Giant Planet Scatterings and Collisions]{Giant Planet Scatterings and Collisions: Hydrodynamics, Merger-Ejection Branching Ratio, and Properties of the Remnants}

\author[Jiaru Li et al.]{
Jiaru Li$^{1,3}$\thanks{E-mail:jiaru\textunderscore li@astro.cornell.edu}, Dong Lai$^{1}$, Kassandra R. Anderson$^{2}$, Bonan Pu$^{1}$
\\
$^{1}$Center for Astrophysics and Planetary Science,
Department of Astronomy, Cornell University, Ithaca, NY 14853, USA\\
$^{2}$Department of Astrophysical Sciences, Princeton University, Princeton, NJ 08544, USA\\
$^{3}$Theoretical Division, Los Alamos National Laboratory, Los Alamos, NM 87545, USA
}

\date{Accepted XXX. Received YYY; in original form ZZZ}

\pubyear{2020}

\begin{document}
\label{firstpage}
\pagerange{\pageref{firstpage}--\pageref{lastpage}}
\maketitle

\begin{abstract}
Planetary systems with sufficiently small orbital spacings can experience planetary mergers and ejections. The branching ratio of mergers vs ejections depends sensitively on the treatment of planetary close encounters. Previous works have adopted a simple ``sticky-sphere'' prescription, whose validity is questionable. We apply both smoothed particle hydrodynamics and $N$-body integrations to investigate the fluid effects in close encounters between gas giants and the long-term evolution of closely-packed planetary systems.  Focusing on parabolic encounters between Jupiter-like planets with $\MJ$ and $2\MJ$, we find that quick mergers occur when the impact parameter $\rp$ (the pericenter separation between the planets) is less than $2\RJ$, and the merger conserved 97\% of the initial mass. Strong tidal effects can affect the ``binary-planet'' orbit when $\rp$ is between $2\RJ$ and $4\RJ$. We quantify these effects using a set of fitting formulae that can be implemented in $N$-body codes.  We run a suite of $N$-body simulations with and without the formulae for systems of two giant planets initially in unstable, nearly circular coplanar orbits. The fluid (tidal) effects significantly increase the branching ratio of planetary mergers relative to ejections by doubling the effective collision radius. While the fluid effects do not change the distributions of semi-major axis and eccentricity of each type of remnant planets (mergers vs surviving planets in ejections), the overall orbital properties of planet scattering remnants are strongly affected due to the change in the branching ratio. We also find that the merger products have broad distributions of spin magnitudes and obliquities.
\end{abstract}


\section{Introduction}
\label{sec:intro}
A system of two or more planets on nearly circular, coplanar orbits can be dynamically unstable if the planet spacing is too small \citep[e.g,][]{Gladman1993,Chambers1996,Zhou2007,Smith2009,Funk2010,Deck2013,Petit2018}. The instability results in strong scatterings or close encounters between planets, leading to violent outcomes such as planetary mergers and ejections. Since the early days of exoplanet detection, the importance of strong planet scatterings in shaping the architecture of planetary systems has been recognized \citep{Rasio1996,Weidenschilling1996,Lin1997}. Indeed, there now exists a large literature on giant planet scatterings \citep{Ford2001,Adams2003,Chatterjee2008,Ford2008,Juric2008,Nagasawa2011,Petrovich2014,Frelikh2019,Anderson2020}. These works typically apply a large number of $N$-body simulations to different initial conditions to investigate the scattering outcomes in a statistical manner. Some are notable for their attempts to reproduce the exoplanetary eccentricity distribution for a wide range of initial conditions \citep{Ford2008,Juric2008,Anderson2020}.

The branching ratio, referring to the the probability of planet collisions/mergers vs ejections in planetary scattering outcomes, is a crucial factor in determining the overall eccentricity distribution, as collisions are much less efficient at producing large eccentricities \citep{Ford2008,Juric2008,Anderson2020}. To derive the branching ratio from $N$-body simulations, a prescription for planet collisions is needed.  Previous works have either neglected planet collisions or adopted the so-called ``sticky-sphere'' approximation to handle close encounters between planets. This approximation assigns a radius, usually the physical radius of the planet, to each point mass in the simulation. When the separation between two point masses is less than the sum of their radii, the two masses immediately merge into a single object in a manner that conserves mass, momentum, and the position of the center of mass.

Several assumptions in the ``sticky-sphere'' approximation are questionable. For example, the merger prescription in this approximation overlooks the detailed kinematics of the collision but instead assumes all planet collisions are the same. However, previous studies have shown that the outcomes of collisions with different kinematics can substantially diverge \citep{Agnor2004,Asphaug2006,Leinhardt2011,Stewart2012,Burger2019,Emsenhuber2019a}. Another problematic situation is when the planets do not collide but bypass each other with their minimum separation comparable to their radii. For such close enconters, the planets can distort each other through tidal effects and cause energy dissipation or even mass transfer. After all, there is no rigorous justification as to why a merger should happen if and only if two planets touch each other.

The issues discussed above have sometimes been recognized, but are usually ``swept under the rug'' in published papers. Addressing these issues requires hydrodynamics simulations of close encounters between planets. Current hydrodynamics simulations on this topic mostly apply to planetesimal collisions or late bombardment process, during which collisions could be very hyperbolic and the reaccretion efficiency is uncertain \citep{Leinhardt2011,Stewart2012}. With a few exceptions \citep{Hwang2017,Hwang2018}, tidal interactions between planets and their effects on the scattering outcomes have not been investigated systematically to date.

In this work, we hope to address the aspects that are missing from current studies. We carry out fluid simulations of close encounters between two giant planets that approach each other in a parabolic orbit. We study the conditions for the two planets to merge and the properties of the merger products. Our hydrodynamics simulations also quantify how much the planets' trajectories are modified during a ``bypassing'' encounter. We then apply our hydrodynamics simulation results (including fitting formulae) in long-term orbital integrations of scatterings of two giant planets. We determine how the fluid effects influence the outcomes of the scatterings, including the merger/ejection branching ratios, the orbital property of the surviving planets, and the spin property of the merger remnants.

The rest of this paper is organized as follows. In Section~\ref{sec:fluid}, we use smoothed-particle hydrodynamics to simulate close encounters between two giant planets and analyze the results. In Section~\ref{sec:prescription}, we present a close encounter prescription to be used for $N$-body codes based on the results from Section~\ref{sec:fluid}. In Section~\ref{sec:nbody}, we run a large set of orbital integrations of systems with two planets, both with and without the prescription derived in Section~\ref{sec:prescription}, which allows us to determine the significance of the fluid effects for the long-term evolution of the planetary systems. We present our conclusions in Section~\ref{sec:conclusion}.

\section{Hydrodynamics Simulations of Encountering Gas Giants}
\label{sec:fluid}

\subsection{Simulation setup}
\label{sec:fluid-init}

We perform simulations of gas giant encounters using the smoothed particle hydrodynamics (SPH) code \textsc{ StarSmasher}\footnote{\textsc{StarSmasher} is available at \url{https://jalombar.github.io/starsmasher/}} \citep{Rasio1991,Lombardi1999,Faber2000,Gaburov2010,Gaburov2018}. \textsc{StarSmasher} balances the accuracy and speed by using NVIDIA graphics cards to calculate the gas self-gravity through a direct summation of each pairwise gravitational interaction between SPH particles \citep{Gaburov2010a,Gaburov2010}.

In this work, we consider two gas giants with masses \footnote{Throughout this paper, the subscript ``J'' specifies the corresponding Jovian value.} $M_1 = 2\MJ$, $M_2=1\MJ$ and radii $R_1=R_2=\RJ$. The two planets are initialized and relaxed in isolation. We construct each planet by placing $10^5$ SPH particles uniformly inside a sphere, and assigning them masses ($m_i$) and specific internal energies ($u_i$) according to the equation of state $P = K\rho^{\gamma}$, where $\gamma = 1+\frac{1}{n}$. We use $n=1$ to model gas giants (see \citealt{Guillot2005} and \citealt{Guillot2014} for justification). After initialization, we switch to the more general equation of state, $P = (\gamma - 1)\rho u$, and relax them until the total kinetic energy of the particles (in the rest frame of each planet) diminishes to less than $10^{-5}$ of the total binding energy.

The dynamical simulations are done in the center-of-mass frame of the two encountering planets. Most gas giant encounters occur in parabolic relative trajectories \citep{Anderson2020}. Hence, we set the two relaxed planets in an initial condition such that:
\begin{enumerate}
    \item Their centers of mass, ${\bf x}_{\rm CM}^{(1)}$ and ${\bf x}_{\rm CM}^{(2)}$, are $15\RJ$ away from each other.
    \item They have the initial velocities, ${\bf v}_{\rm CM}^{(1)}$ and ${\bf v}_{\rm CM}^{(2)}$, such that, if they were point masses, their relative trajectory would be a parabola ($e=1.0$) with a specified pericenter distance $\rp$.
\end{enumerate}
The only free parameter in this set-up is $\rp$. We run simulations for 20 different $\rp$'s that spread equally from $0.2\RJ$ to $4.0\RJ$. Every dynamical run includes at least one pericenter passage for the planetary binary. We run the simulations until the post-encounter products settle down appropriately (see below).

\subsection{Identify the post-encounter products}
\label{sec:fluid-grouping}

Since the SPH code we use does not distinguish the fluid particles belonging to different planetary bodies, we must identify a set of the SPH particles, denoted as $S_k$, that can be treated as a coherent body (``planet $k$''). The mass, position, and velocity of each post-encounter planet are
\begin{equation}
\begin{split}
    \Mk & = \sum_{i \in S_k} m_i, \\
    \xCMk & = \frac{1}{\Mk}\sum_{i \in S_k} m_i {\bf x}_i,\\
    \vCMk & = \frac{1}{\Mk}\sum_{i \in S_k} m_i {\bf v}_i,
\end{split}
\end{equation}
where $m_i$, ${\bf x}_i$, and ${\bf v}_i$ are the mass, position, and velocity of a SPH particle $i$. Particles in the same set do not necessarily form a sphere. For convenience, we define the radius of a post-encounter planet as the radius of an imaginary sphere that contains $90\%$ of the planetary mass, i.e.
\begin{equation}
    R^{(k)} = \min(\{ R | \Sigma_{i \in S_k}^{| {\bf x}_i - \xCMk| < R} m_i \leq 0.9 \Mk \} ).
\end{equation}
This is sometimes called the Lagrange radius.

Our method to identify post-encounter planets utilizes the Bernoulli constant. Each SPH particle $i$, if belonging to planet $k$, has a specific enthalpy
\begin{equation}
    h_{i}^{(k)} = \frac{1}{2}({\bf v}_i - \vCMk)^2 + u_i + \frac{P_i}{\rho_i} + \phi_i^{(k)},
\end{equation}
where $u_i$ and $\rho_i$ are its specific internal energy and density, $P_i$ is the local pressure, and $\phi_i^{(k)}$ is the gravitational potential due to the gas in $S_k$. Along a streamline with no dissipation, $h_i$ should be a constant according to the Bernoulli theorem. The objective is to minimize the sum
\begin{equation}
\begin{split}
    H = \sum_{i \in S_1}m_i h_i^{(1)} + \sum_{i \in S_2}m_i h_i^{(2)}.
\end{split}
\end{equation}
Particles with $h_{i}^{(k)} > 0$ for both $k = 1$ and $2$ do not belong to $S_1$ nor $S_2$ --  they are considered as the ejecta, unbound from the system. Note that in the case of a merger, $H$ is minimized when $S_2$ is a `null' set containing a negligible amount of particles.

For numerical efficiency, we do not attempt to find the global minimum of $H$ by checking all of $3^N$ possible groupings, where $N$ is the total number of SPH particles. Instead, similar to  \cite{Emsenhuber2019b}, we use a Friends-of-Friends (FoF) method to obtain an initial guess first. The FoF algorithm grows a remnant planet by gluing neighboring particles to a cluster when certain conditions are satisfied. In our implementation, we employ the most massive SPH particle as the seed and then grow a planet (cluster of particles) iteratively. At each growing step, every SPH particle $i$ inside a cluster will search for neighboring particles $j$'s that are outside of the cluster but within two times its SPH smoothing length $h_i$ from its location. If a neighboring particle has density $\rho_j > 10^{-3}\MJ/\RJ^3$, we will then add it to the cluster as a new member. Once this planet finds no qualified neighbors, we use the most massive unclustered SPH particle to grow the second planet. A $k$-d tree method \citep{Kennel2004} is used to search for neighbors. After this FoF procedure, we minimize $H$ by iteratively updating the grouping using gradient descent.

\subsection{Results}
\label{sec:fluid-res}

As noted in Section~\ref{sec:intro}, the standard way of resolving close encounters in $N$-body planetary integrators is to use the `sticky-sphere' approximation. It handles close encounters following three rules: (1) the two planets merge if and only if they physically touch each other; (2) mergers always conserve the mass and momentum of the planetary binary; (3) when there is no physical collision, the encounter is equivalent to the gravitational interaction between point masses. We examine the three assumptions in the three subsections below.

\subsubsection{Merger conditions}

\begin{figure*}
	\includegraphics[width=1.8\columnwidth]{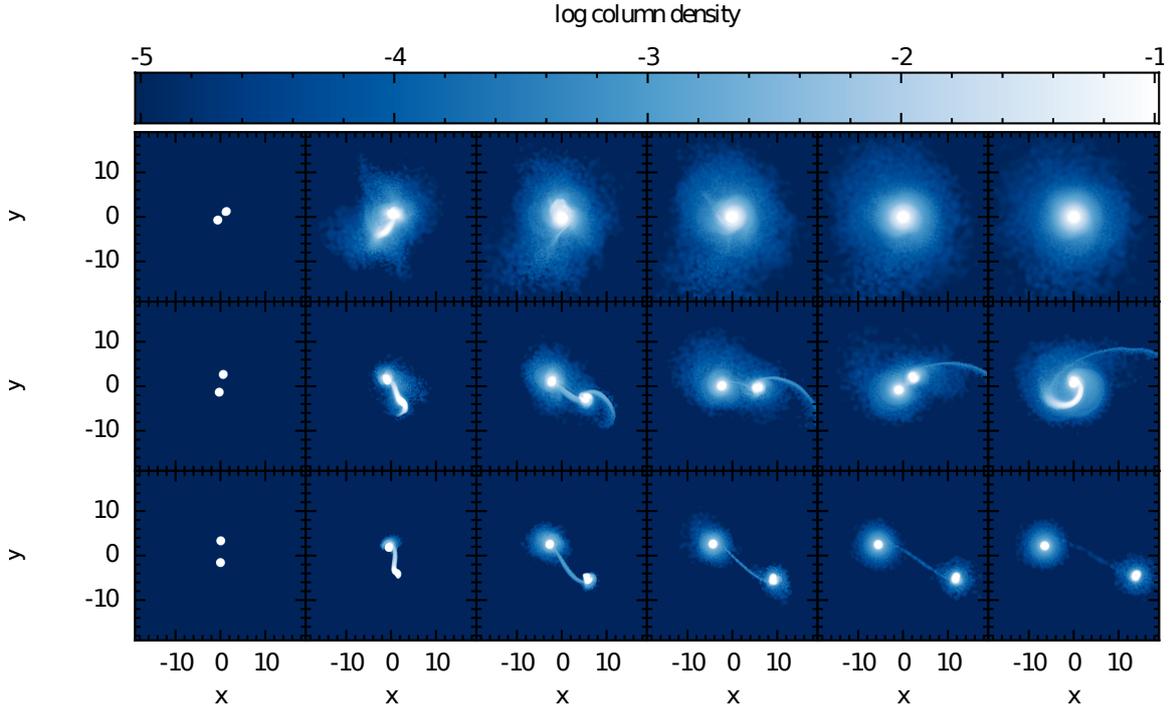}
    \caption{Three types of outcomes for two-planet encounters. Each panel shows the gas column density (in units of $\MJ/\RJ^2$) from the SPH simulations. The three rows are simulations with $\rp = 0.4, 1.6, 2.4 \RJ$ from top to bottom. They correspond to the one-shot merger, two-step merger, and the bypassing scenarios, respectively. From left to right, the columns are the snapshots at time steps $t = 15, 25, 35, 45, 55, 65 \sqrt{\RJ^3/G\MJ}$ since the start of each simulation. This plot is produced using \textsc{splash} \citep{Price2007}.}
    \label{fig:fluid-merger_movie}
\end{figure*}

The outcomes of two-planet encounters can be divided into three categories. Fig.~\ref{fig:fluid-merger_movie} shows an example of the time evolution for each type of outcome:
\begin{enumerate}
    \item One-shot merger (the top row of Fig.~\ref{fig:fluid-merger_movie}): the two planets collide nearly head-on and merge immediately.
    \item Two-step merger (the middle row of Fig.~\ref{fig:fluid-merger_movie}): the planets experience two consecutive close encounters, where the second encounter leads to a complete tidal disruption of $M_2$ (the lower-mass planet). The disrupted mass orbits around and accretes onto $M_1$ (the higher-mass planet), effectively leading to a merger.
    \item Bypassing (the bottom row of Fig.~\ref{fig:fluid-merger_movie}): the two planets survive the first encounter and do not come back together for a second encounter for an extended period of time.
\end{enumerate}

\begin{figure}
    \includegraphics[width=\columnwidth]{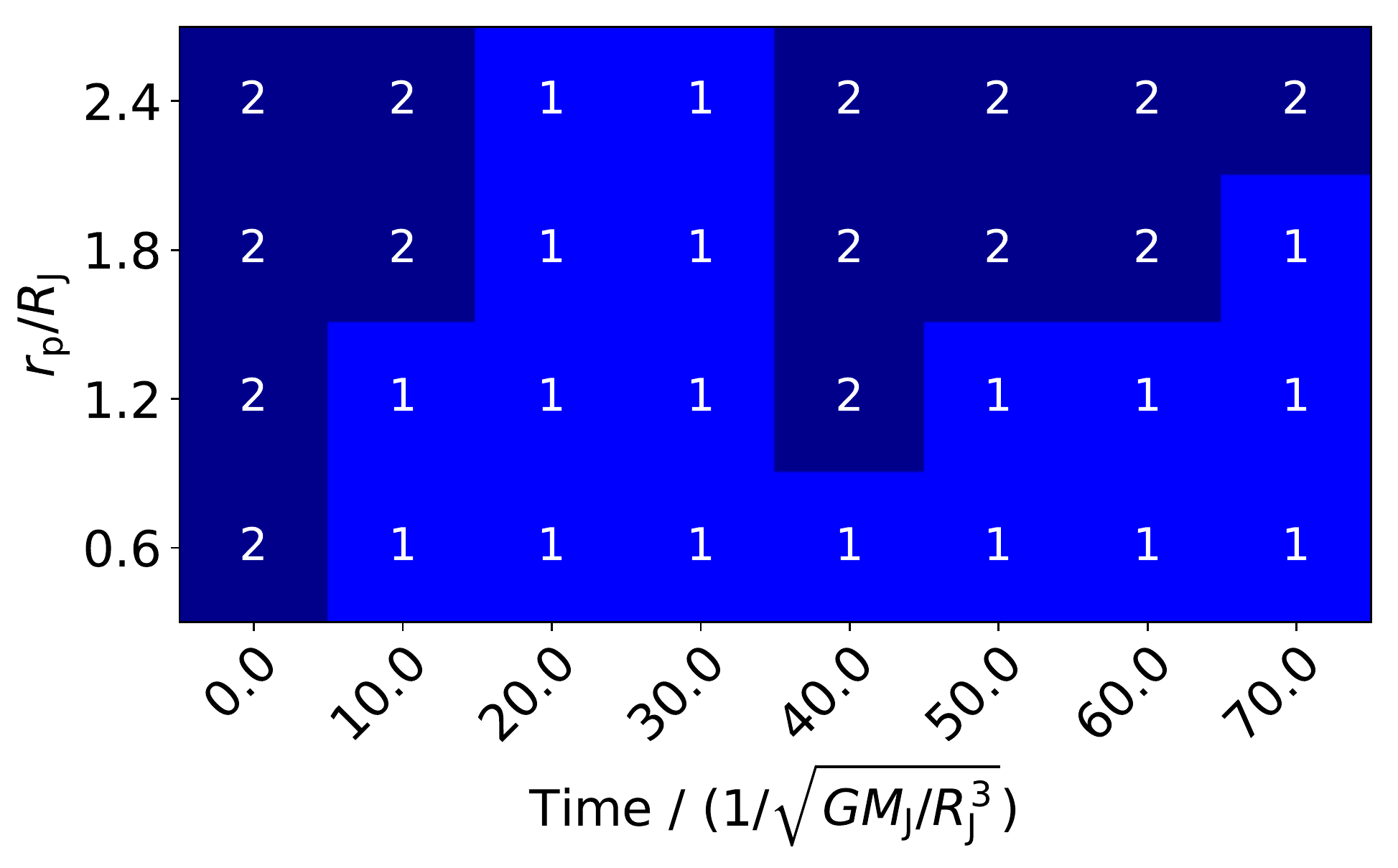}
    \caption{Time evolution of the number of ``planets'' (with mass $>0.1\MJ$) in various numerical hydrodynamics simulations. A number `1' implies that there is only one planet at the end of the corresponding time bin. A number `2' implies that two planets are distinguishable by our planet grouping algorithm.}
\label{fig:fluid-merger_test}
\end{figure}

In addition to visual inspections, we also use the grouping algorithm described in Section~\ref{sec:fluid-grouping} to determine the number of remaining planets by examining the mass binning. Fig.~\ref{fig:fluid-merger_test} shows this result. (i) Encounters with $\rp<\RJ$ are nearly head-on and always lead to one-shot mergers. (ii) When $ \RJ < \rp < R_1+R_2 = 2\RJ$, the angular momentum in the planetary binary system prevents an immediate merger. However, after losing orbital energy at the first collision, the planets can loop back rather quickly and the less massive planet becomes vulnerable to tidal disruption at the second encounter. The gas then form a single planetary body through accretion. (iii) Encounters with $\rp > R_1+R_2 = 2\RJ$ can recover from a ``fuzzy'' period of pericenter passage and will not come across a second encounter before the end of our simulations, which is roughly 50 units of time ($\sqrt{\RJ^3/G\MJ}$) after the first encounter.

We note that, if we keep running these systems, all of the bypassing binary can loop back for second encounters. Hence, the exact boundary between two-step merger events and bypassing events does not exist. We argue in Section~\ref{sec:fluid-dis-rp} that $\rp=2\RJ$ is the optimal choice for this boundary in $N$-body simulations.

In summary, although some mergers require two steps, the condition for mergers is the same as the first rule used in standard $N$-body simulations, i.e. the planets must touch each other along their point-mass trajectories.

\subsubsection{Merger products}
\label{sec:fluid-res-merger}

Table~\ref{tab:fluid-merger_conserve} shows the properties of the merger products. The mergers preserve at least $97\%$ of the total mass from the two colliding planets. In the initial center-of-mass frame of the two planets, the merger products barely gain any velocity due to collision-induced mass loss, especially at small $\rp$. These two results suggest that mergers can be treated as perfect inelastic collisions which conserve the mass and momentum of the planetary binaries.

\begin{table*}
	\centering
    \caption{The properties of the merger product at $t = 150\sqrt{\RJ^3/G\MJ}$ in hydrodynamics simulations with different $r_p$. The mass is determined using the method described in Section~\ref{sec:fluid-grouping}. The speed (in units of $\sqrt{GM_J/R_J}$) is the bulk speed of all the bounded SPH particles. The spin angular momentum of the merger product is in units of $L_{\rm in}$ from equation~(\ref{eq:Lin}). The final radius is measured as the Lagrange radius (the spherical radius that encloses $90\%$ planetary mass). The initial Lagrange radius of each planet is $0.85\RJ$ for the $n=1$ polytrope density profile adopted in this study.}
    \label{tab:fluid-merger_conserve}
    \begin{tabular}{ lccccccccc } 
     \hline
     $\rp  (\RJ)$   & 0.2   & 0.4   & 0.6   & 0.8   & 1.0
     & 1.2  & 1.4   & 1.6   & 1.8\\
     \hline
     Mass $(\MJ)$   & 2.98  & 2.99  & 2.98  & 2.99  & 2.98
     & 2.97 & 2.95  & 2.94  & 2.92 \\ 
     Speed $(\sqrt{G\MJ/\RJ})$ & <0.001  & <0.001  & 0.001  & <0.001  & 0.002 
     & 0.003 & 0.007  & 0.008  & 0.012\\ 
     Spin $(L_{\rm in})$  & 0.99  & 1.00  & 0.99  & 0.99  & 0.97 
     & 0.95 & 0.89  & 0.85  & 0.80\\ 
     Radius $(0.85\RJ)$             & 3.37  & 3.70  & 3.48  & 3.91  & 4.55
     & 4.32 & 4.57  & 5.00  & 4.68 \\ 
     \hline
    \end{tabular}
\end{table*}

The merger products are fast-spinning due to the angular momentum of the incidental binary orbit,
\begin{equation}
\label{eq:Lin}
    L_{\rm in} = \frac{M_1M_2}{M_1+M_2}\sqrt{2G(M_1+M_2)\rp}.
\end{equation}
More than $95\%$ of the orbital angular momentum are inherited by the merged object when $\rp \leq 1.2\RJ$. Mergers with $\rp>1.2\RJ$ conserve $80\%$ to $95\%$ of the initial angular momentum. In all cases, the direction of the spin is the same the orbital angular momentum of the incidental binaries. Obviously, the merger products are rather ``inflated'' due to rotational support compared to the initial planets. Long-term evolution of these ``soft'' gas bodies would be of interest, but is beyond the scope of this work.

\subsubsection{Bypassings}
\label{sec:fluid-res-te}

For encounters with $\rp>R_1+R_2$, the planets bypass each other. When the separation between the planets increases back to several planetary radii, the interaction between the planets become point-mass-like again. Hence, the post-encounter mass of the planets and the orbital elements of their relative motions are well-defined and can be parametrized as functions of $\rp$.

\begin{figure}
    \includegraphics[width=\columnwidth]{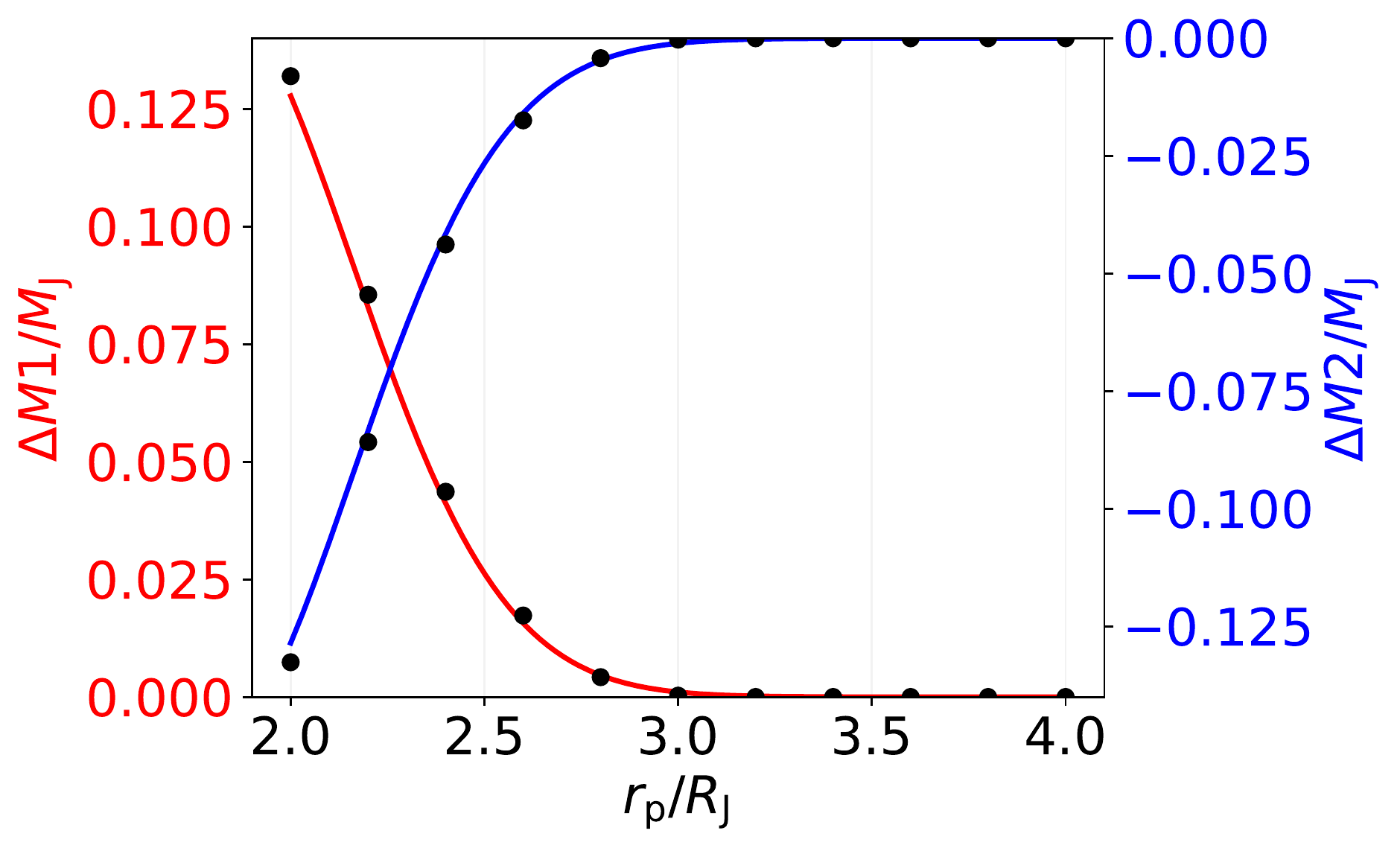}
    \caption{Mass transfer during bypassings. The black dots are data from the SPH simulations and the curves are the fitting formulae (see Table 3). The red line is for the mass gained by $M_1$, while the blue is for the mass loss from $M_2$.}
	\label{fig:fluid-dm}
\end{figure}

\begin{table*}
    \begin{center}
    \caption{The final radii of the two planets after close bypassings. The radius is defined as the Lagrange radius that encloses $90\%$ of planetary mass. The initial planet Lagrange radius is $0.85\RJ$ for $n=1$ polytropes.}
    \label{tab:R}
    \begin{tabular}{ c|ccccccccccc } 
    \hline
    $\rp  (\RJ)$   & 2.0   & 2.2   & 2.4   & 2.6   & 2.8
    & 3.0  & 3.2   & 3.4   & 3.6   & 3.8   & 4.0\\
    \hline
    $R_1$ $(0.85\RJ)$  & 1.10  & 1.05  & 1.02  & 1.01  & 1.00 
    & 1.00 & 1.00  & 1.00  & 1.00  & 1.00  & 1.00\\ 
    $R_2$ $(0.85\RJ)$  & 1.25  & 1.18  & 1.12  & 1.08  & 1.05
    & 1.02 & 1.01  & 1.00  & 1.00  & 1.00  & 1.00\\ 
    \hline
    \end{tabular}
    \end{center}
\end{table*}

Close encounters induce mass transfer from the less massive planet to its companion. Fig.~\ref{fig:fluid-dm} shows the mass exchange between the two planets. The amount of transferred mass increases steeply as $\rp$ becomes smaller than $2.7\RJ$, which is approximately the tidal radius. The fact that mass loss from $M_2$ approximately equals to the gain by $M_1$ implies that the tidal interaction conserves the total mass. Table~\ref{tab:R} displays the changes in planetary radii. Both planets slightly grow in size from the initial Lagrange radii of $R_1 = R_2 =0.85\RJ$. The lighter planet $R_2$ is affected more comparing to $R_1$. However, neither radius changes are significant.

\begin{figure}
	\includegraphics[width=\columnwidth]{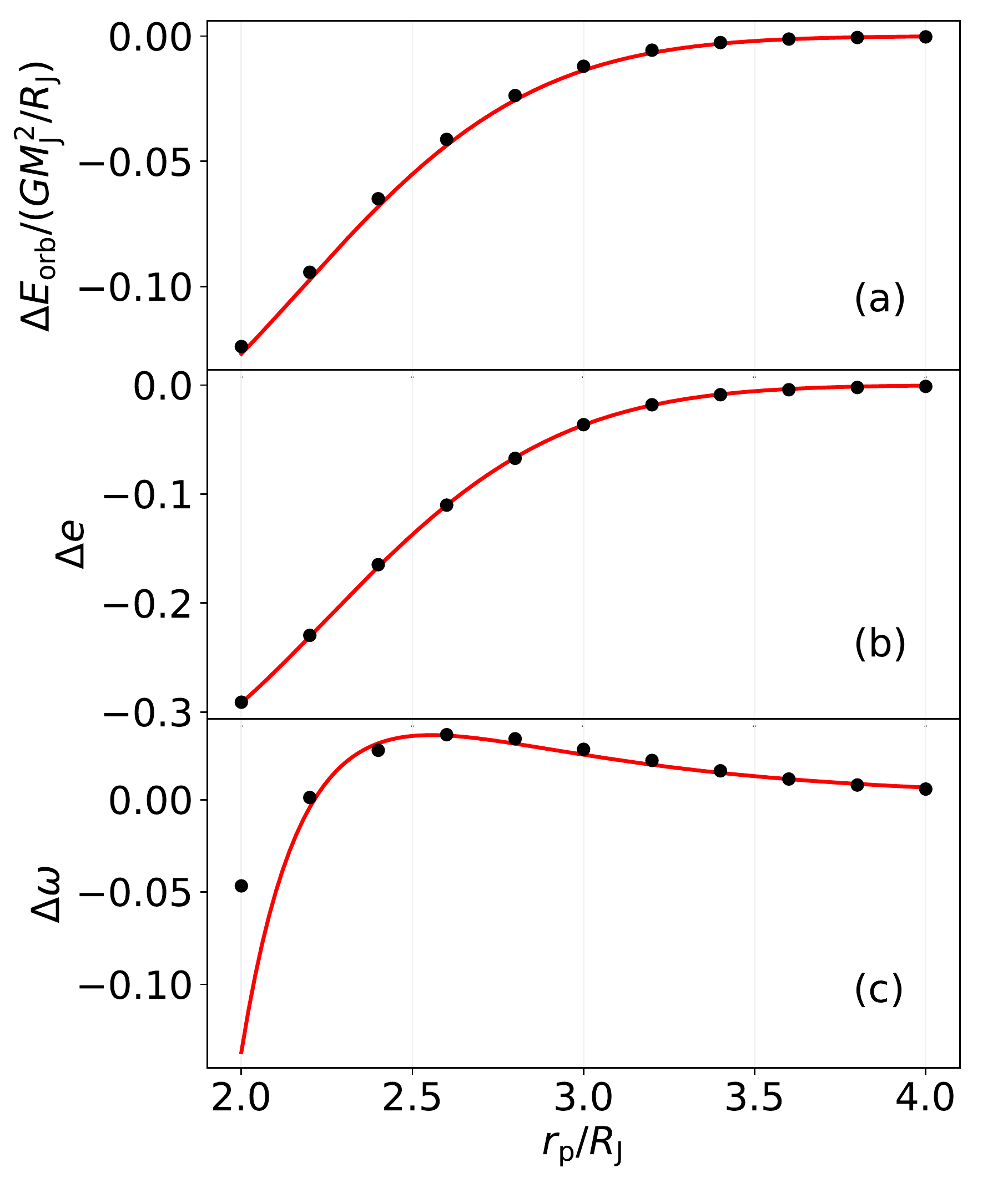}
	\caption{The change of various orbital quantities for encounters with different $\rp$'s. The dots represent data from SPH simulations, and the red curve is the fitting formula (see Table~\ref{tab:fluid-fit}). Panel (a): the orbital energy changes $\Delta E_{\rm orb}$; Panel (b): the eccentricity changes $\Delta e$; Panel (c): the shifts of longitude of pericenter $\Delta\omega$ in radians, with a positive $\Delta \omega$ implies that the pericenter is shifted in the direction of the binary orbit.}
	\label{fig:fluid-dEew}
\end{figure}

The post-encounter binary orbits are different from the incident orbits. Even in a very gentle encounter with no mass exchange, planets can always excite oscillations inside their partners (``dynamical tides''), which cause the binary orbit to lose energy. Draining energy from the orbit also changes the eccentricity of the orbit. Panel (a) and (b) of Fig.~\ref{fig:fluid-dEew} show the changes of orbital energy and eccentricity obtained from our simulations.  The post-encounter orbit is still confined in the initial orbital plane.  However, the direction of the eccentricity vector may change within the orbital planet. We measure the new orbital orientation in terms of the shift of the longitude of pericenter, $\Delta\omega$, from our simulations. Panel (c) of Fig.~\ref{fig:fluid-dEew} shows our result.

In summary, close encounters can induce mass transfer between planets and modify the binary orbits. These changes can be parametrized with the impact pericenter distance, $\rp$, using some fitting formulae. Table~\ref{tab:fluid-fit} presents the formulae based on the results of our simulations.

\begin{table}
    \caption{Fitting formulae for mass transfer, orbital energy change, eccentricity change, and the shift in the longitude of pericenter for encounters of two planets $M_1=2M_J$ and $M_2=M_J$.  The formulae apply only to the non-merger encounters, i.e. those with $\rp \geq R_1 + R_2 = 2\RJ$.}
    \label{tab:fluid-fit}
    \begin{tabular}{ cc|ccccc } 
    \hline
    Fitting Formula & & $\Delta M_1$  & $\Delta M_2$  & $\Delta E_{\rm orb}$  & $\Delta e$    & $\Delta \omega$ \\ 
    \hline
     $Ae^{-b(\rp-c)^2}$ 
    & A & +0.152         & -0.153       & -0.167            & -0.356   & -\\
    & b & +3.303        & +3.283        & +1.116           & +1.128    & -\\
    & c &    +1.771      & +1.770          &  +1.503            &  +1.581   & -\\
     $-a\rp^{-7} + b\rp^{-6}$   
    & a & -             & -             & -                 & -         & +191\\
    & b & -             & -             & -                 & -         & +39\\
    \hline
    \end{tabular}
\end{table}

\subsection{Discussion}

\subsubsection{Two-step mergers}
\label{sec:fluid-dis-rp}

In the above, we label a collision event ``two-step merger'' if the planets quickly experience a second tidal encounter after the first one. Here, we discuss what `quickly' should mean and justify our choice of the boundary between two-step merger and bypassing.

\begin{figure}
    \includegraphics[width=\columnwidth]{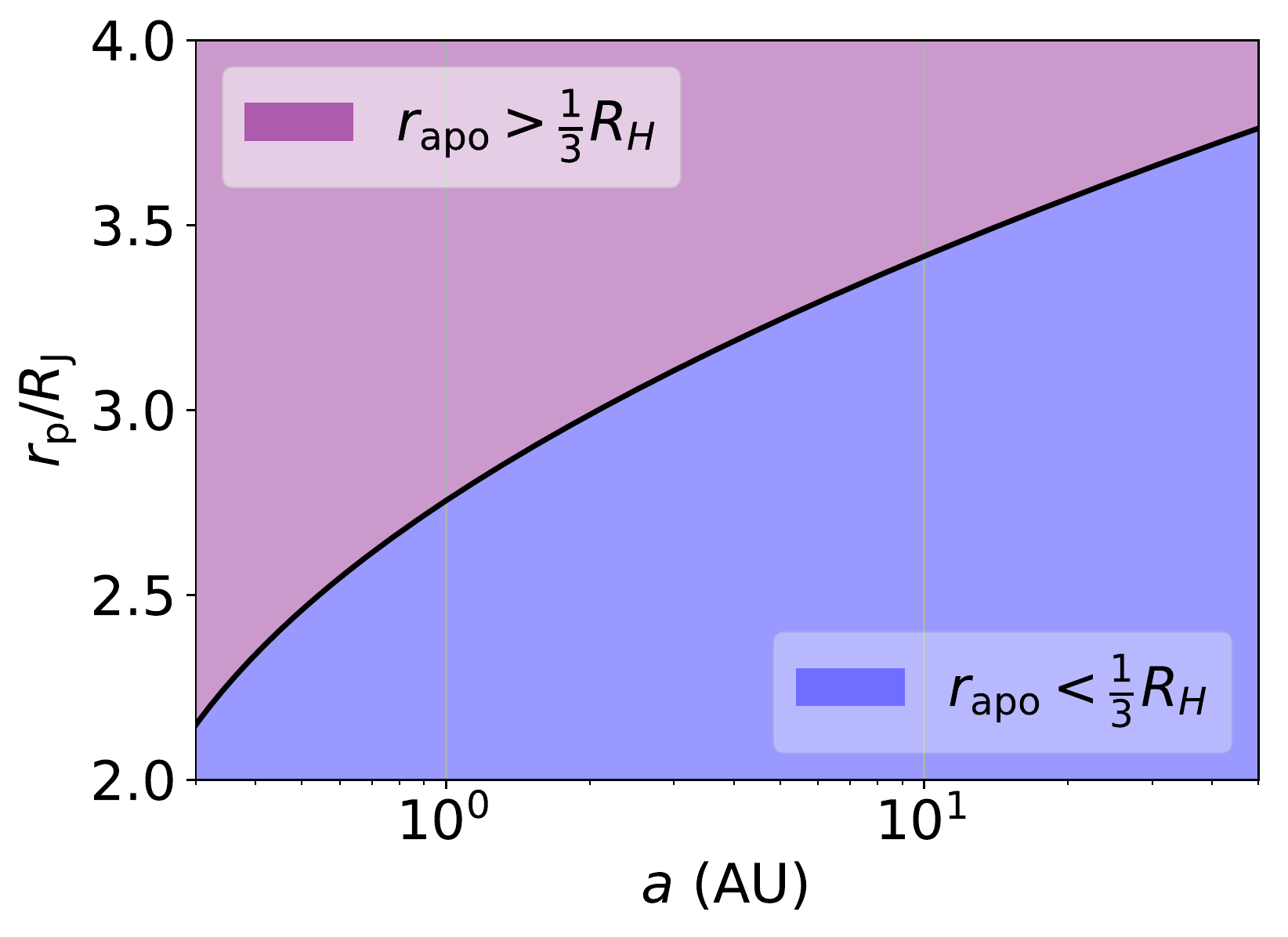}
    \caption{The criterion for the two planets to stay within $1/3$ of the mutual Hill radius from each other after their first close encounter at different distance $a=(a_1+a_2)/2$ from the star. The Hill radius $R_{\rm H}$ is calculated using equation~(\ref{eq:MHR}) and $M_{\star}=M_{\sun}$. The black curve is the condition for $r_{\rm apo} = \frac{1}{3}R_{\rm H}$, where $r_{\rm apo}$ is the post-encounter apocenter distance between the planets. We evaluate $r_{\rm apo}$ with the fitting formulae in Table~\ref{tab:fluid-fit}.}
    \label{fig:fluid-RH}
\end{figure}

First, this second encounter must happen before the tidal gravity from the central star ``disassociates'' the binary planets. Our fluid simulations neglect the influence of the central star. \cite{Emsenhuber2019a} suggest that when the post-encounter apocenter distance is inside roughly $1/3$ of the Hill radius from the primary planet, the loop-back process is not strongly affected by the star. This criterion is more robust for smaller $\rp$. As the post-encounter planet-planet separation increases, the loop-back process becomes increasingly random. Based on our fluid simulation results, the $1/3$ Hill radius is reached at $\rp \approx 2.7\RJ$ when the encounter takes place at $1$AU from a solar-mass star. Fig.~\ref{fig:fluid-RH} shows the values of this critical $\rp$ at different distance ($a$) from the star. For a destructive second encounter to happen, $\rp$ must be less than the critical value. The smaller, the better.

Second, a quick second encounter usually comes before the planets can recover their point-mass properties. To treat a planet as a point mass, it must not only be recognized by our grouping algorithm, but should also have converged mass and orbital elements. We have found that for the simulations with $\rp \approx 1.8\RJ$ or smaller, the post-(first-)encounter masses do not converge with respect to time before the second encounter happens. Collisions with $1.8\RJ < \rp < 2.0\RJ$ are not covered in our suite of simulations, so $\rp \approx 2.0\RJ$ is a cautious estimation of the minimum impact parameter for both planets to recover. 

From the reasons above, we conclude that the two planets merge if their impact pericenter distance is less than $2.0\RJ$. For $\rp\leq2\RJ$, a second encounter is guaranteed. For $\rp\geq2\RJ$, the planets can be treated as point masses. This choice also has the most intuitive physical meanings, i.e., physical collisions lead to mergers.

\subsubsection{Different mass ratios}
\label{sec:fluid-mass-ratio}
We repeat our numerical simulations with two planets of masses of $M_1 = 1.5\MJ$ and $M_2 = 1\MJ$ and find similar results in terms of the merger conditions and the properties of the merger products. The tidal effects between the bypassing planets can be evaluated using the fitting formulae in Table~\ref{tab:fluid-fit-2}. The expressions are the same as the in Table~\ref{tab:fluid-fit}, with different fitting coefficients.

\begin{table}
    \caption{Same as Table 3, but with $M_1=1.5\MJ$ and $M_2=1\MJ$.}
    \label{tab:fluid-fit-2}
    \begin{tabular}{ cc|ccccc } 
    \hline
    Fitting Formula &   & $\Delta M_1$  & $\Delta M_2$  & $\Delta E_{\rm orb}$  & $\Delta e$    & $\Delta \omega$ \\ 
    \hline
    $Ae^{-b(\rp-c)^2}$ 
    & A & +0.080         & -0.0808       & -0.268            & -0.647   & -\\
    & b & +5.699        & +5.715        & +1.329           & +1.283    & -\\
    & c & +1.871          & +1.870         &  +1.267            &  +1.355   & -\\
    $-a\rp^{-7} + b\rp^{-6}$   
    & a & -             & -             & -                 & -         & 145\\
    & b & -             & -             & -                 & -         & 67\\
    \hline
    \end{tabular}
\end{table}

\section{An improved Prescription for Close Encounters in $N$-Body simulations}
\label{sec:prescription}

Based on the results of Section~\ref{sec:fluid}, we suggest the following prescription for treating close planetary encounters in $N$-body simulations: Suppose two planets approach each other on a point-mass trajectory with the closest separation $r_p$, 
\begin{enumerate}
    \item If $\rp<2\RJ=R_1+R_2$, the planets merge in a manner that conserves the total mass and momentum.
    \item If $\rp$ is between $2\RJ$ and $4\RJ$, the planets exit the encounter in a new relative trajectory given by the fitting formulae in Tables~\ref{tab:fluid-fit} and \ref{tab:fluid-fit-2}.
    \item If $\rp>4\RJ$, the hydrodynamical effects are small and no modification is needed.   
\end{enumerate}

\begin{figure}
    \includegraphics[width=\columnwidth]{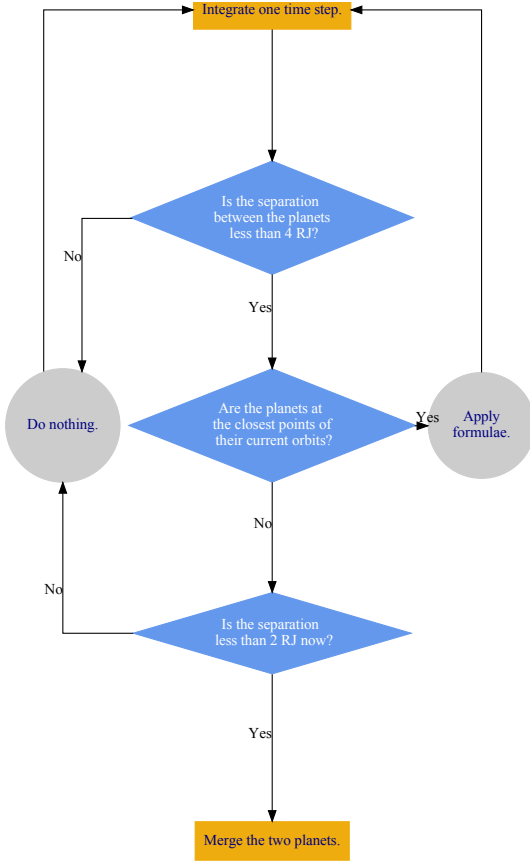}
    \caption{Implementation of close planetary encounters in $N$-body simulations.  After each time step of integration, we may want to merge the two planets, apply the fitting formulae, or do nothing until the next time step. This decision is made by the code based the answers to three true-or-false questions: (i) whether the separation between the two planets is less than $4\RJ$, (ii) whether the planets are at the minimum separation of their current trajectories, and (iii) whether they have already hit each other. The code should follow the order of the steps indicated by the arrows.}
    \label{fig:pres-flow}
\end{figure}

One way to implement the above prescription is to use the impulse approximation as illustrated in Fig.~\ref{fig:pres-flow}.
Before using any fitting formula, the $N$-body simulation should be paused when the binary planets are at their pericenter. From this paused frame, we read the masses of the binary planets, $M_1$ and $M_2$, the position and velocity of the binary center of mass, ${\bf x}_{\rm CM}$ and $\bf{v}_{\rm CM}$, and the relative position and velocity of the two planets, ${\bf r}_{\rm p} = {\bf x}_2 - {\bf x}_1$ and ${\bf v}_{\rm p} = {\bf v}_2 - {\bf v}_1$. The parameter for the fitting formulae, $\rp$, is given by $|{\bf r}_{\rm p}|$. The orientation of the orbital plane , $\hat{\bf n} = \hat{\bf r}_{\rm p} \times \hat{\bf v}_{\rm p}$, will remain the same after applying
the hydrodynamical corrections.

Now the post-encounter planet masses, binary orbital energy, eccentricity, and angular shift of the pericenter ($M_1'$, $M_2'$, $E'_{\rm orb}$, $e'$, and $\Delta\omega$) can be obtained using the formulae in Tables~\ref{tab:fluid-fit} and \ref{tab:fluid-fit-2}. The new relative position and velocity can be calculated as
\begin{equation}
    \begin{split}
    r' & = -\frac{GM'_1 M'_2}{2E'_{\rm orb}} (1-e'), \\
    v' & = \sqrt{2\frac{M'_1 + M'_2}{M'_1M'_2}\left(E'_{\rm orb} + \frac{GM'_1M'_2}{r'}\right)},
    \end{split}
\end{equation}
and the vectors are
\begin{equation}
    \begin{split}
    {\bf r}' & = r' {\bf R}(\hat{\bf n},\Delta\omega)\hat{\bf r}_{\rm p}, \\
    {\bf v}' & = v' {\bf R}(\hat{\bf n},\Delta\omega)(\hat{\bf n} \times \hat{\bf r}_{\rm p}), \\
    \end{split}
\end{equation}
in the binary center-of-mass frame, where ${\bf R}$ is the rotation operator. Hence, the updated positions and velocities are
\begin{equation}
    \begin{split}
        {\bf x}'_1 = -\frac{M'_2}{M'_1+M'_2}{\bf r}' + {\bf x}_{\rm CM}, \qquad & {\bf x}'_2 = +\frac{M'_1}{M'_1+M'_2}{\bf r}' + {\bf x}_{\rm CM}, \\  {\bf v}'_1 = -\frac{M'_2}{M'_1+M'_2}{\bf v}' + {\bf v}_{\rm CM}, \qquad & {\bf v}'_2 = +\frac{M'_1}{M'_1+M'_2}{\bf v}' + {\bf v}_{\rm CM}, \\
    \end{split}
\end{equation}
in the code frame of the $N$-body simulation. This is the full prescription to handle close encounters. It only requires the data that are easily accessible from a $N$-body code ($M_1$, $M_2$, ${\bf x}_{\rm CM}$, ${\bf v}_{\rm CM}$, ${\bf r}_{\rm p}$, and ${\bf v}_{\rm p}$) and returns the updated data that a $N$-body code needs ($M'_1$, $M'_2$, ${\bf x}'_1$, ${\bf x}'_2$, ${\bf v}'_1$, and ${\bf v}'_2$).

\section{Two-Planet Scattering Numerical Experiments}
\label{sec:nbody}

In this section, we carry out simulations of two-planet scatterings using our prescription of planet collisions described in Section~\ref{sec:prescription}. We also compare our results with those using the standard ``sticky-sphere'' presciprtion.

\begin{figure*}
    \includegraphics[width=2\columnwidth]{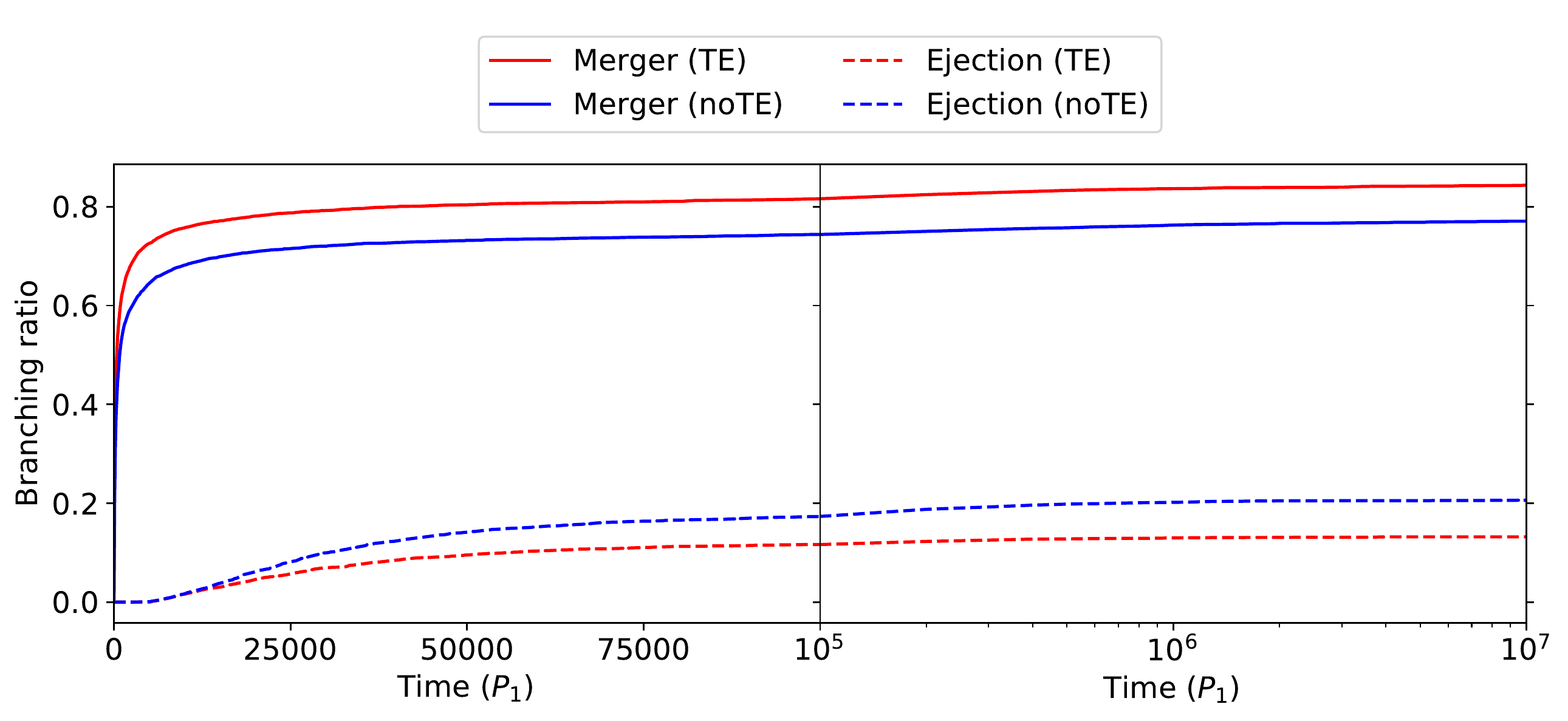}
    \caption{Fraction of systems that have experienced mergers (solid lines) and ejections (dashed lines) as a function of time for the fiducial set of simulations. The red and blue lines represent the results from \texttt{TE} (including ``Tidal Encounter'' prescription) and \texttt{noTE}, respectively. \textit{Left}: The first $10^5$ initial orbital periods of the inner planet, where most of the systems go unstable. \textit{Right}: The evolution after the first $10^5$ initial orbital periods of the inner planet. At $t=10^7P_1$, only $2\%$ of the systems are still stable, and the merger and ejection fractions are $83\%$ and $13\%$ for the \texttt{TE} runs (with the corresponding fractions $75\%$ and $21\%$ for the \texttt{noTE} runs); planet-star collisions occur in the remaining $2\%$ of the systems.}
    \label{fig:nbody-fid-br}
\end{figure*}

\subsection{Setup and fiducial parameters}
\label{sec:nbody-setup}

We perform two-planet scattering experiments using \textsc{rebound}\footnote{\textsc{Rebound} is freely available at \url{http://github.com/hannorein/rebound}.} \citep{Rein2012} with the \textsc{IAS15} integrator \citep{Rein2014}. The close-encounter prescription is implemented as a \textsc{python} function that can be called from \textsc{rebound}. We run a group of simulations using the prescription in Section~\ref{sec:prescription} (following Fig.~\ref{fig:pres-flow}) and another group of simulations using the standard ``sticky-sphere'' prescription.  They are referred to as \texttt{TE} (``Tidal Effects'') and \texttt{noTE}, respectively, since the key difference is whether tidal effects are included for the bypassing planets. We stop a simulation whenever one of the following conditions is reached:
\begin{itemize}
  \item Merger: The separation of the planets is equal to the sum of their physical radii.
  \item Ejection: One of the planets reaches a distance of $1000$~AU from the system's center of mass.
  \item Star-Grazing: The distance between the star and one of the planets is less than the solar radius.
  \item Stability: The integration reaches a chosen time limit without triggering any of the three conditions listed above.
\end{itemize} 
The simulation results are also assorted into the four categories in the ending conditions. 

We begin with a system of two planets with masses $m_1 = 2\MJ$, $m_2 = 1\MJ$ and radii $R_1 = R_2 = \RJ$, orbiting a host star with mass $M_{\star} = M_{\sun}$ and radius $R_{\sun}$. The initial spacing of the planets is set by 
\begin{equation}
    \label{eq:KRH}
    a_2 - a_1 = K R_{\rm H,mut},
\end{equation}
where 
\begin{equation}
    \label{eq:MHR}
    R_{\rm H,mut} = \frac{a_2+a_1}{2}\left(\frac{m_1+m_2}{3M_{\star}}\right)^{1/3},
\end{equation}
is the mutual Hill radius. For each planet, we sample the initial eccentricity in the range $[0.01,0.05]$, the initial inclination from $[0^{\circ},2^{\circ}]$, and the argument of pericenter, longitude of ascending node, and mean anomaly in the range $[0,2\pi]$, assuming they all have uniform distributions.

Our fiducial set of simulations consists of 5000 \texttt{TE} runs and 5000 \texttt{noTE} runs with $a_1=1$AU and $K=2.5$. The integration time limit is set to $10^7P_1$, where $P_1$ is the initial orbital period of the inner planet. The results from the fiducial runs are presented in Sections~\ref{sec:nbody-fid}. In Sections~\ref{sec:a1} and~\ref{sec:K}, we investigate how the results depend on the initial $a_1$ and $K$.

\subsection{Fiducial results}
\label{sec:nbody-fid}

\subsubsection{Branching ratios}
\label{sec:nbody-fid-br}

Fig.~\ref{fig:nbody-fid-br} shows the time evolution of the branching ratio in the fiducial runs. Since $K=2.5$ does not satisfy the criterion for Hill Stability ($K>2\sqrt{3}$; see \citealt{Gladman1993}), most of the systems quickly go unstable (only about $2\%$ of the systems remain stable for $10^7P_1$). The branching ratios converge after $10^6P_1$. The merger of planets is the most common the outcome: $75.2\%$ of the \texttt{noTE} runs and $83.3\%$ of the \texttt{TE} runs end in this way, and most of these mergers happen within $10^4P_1$ from the beginning of the simulations. The ejection of the low-mass planet $M_2$ constitutes $20.1\%$ of the \texttt{noTE} runs and $13.0\%$ of the \texttt{TE} runs, and most of the ejections finish within $10^5P_1$. Planet-star collision and the ejection of $M_1$ together contribute about $2\%$ of the outcomes. For simplicity, from this point forward, we consider only the ejection of $M_2$.

\begin{figure}
    \includegraphics[width=\columnwidth]{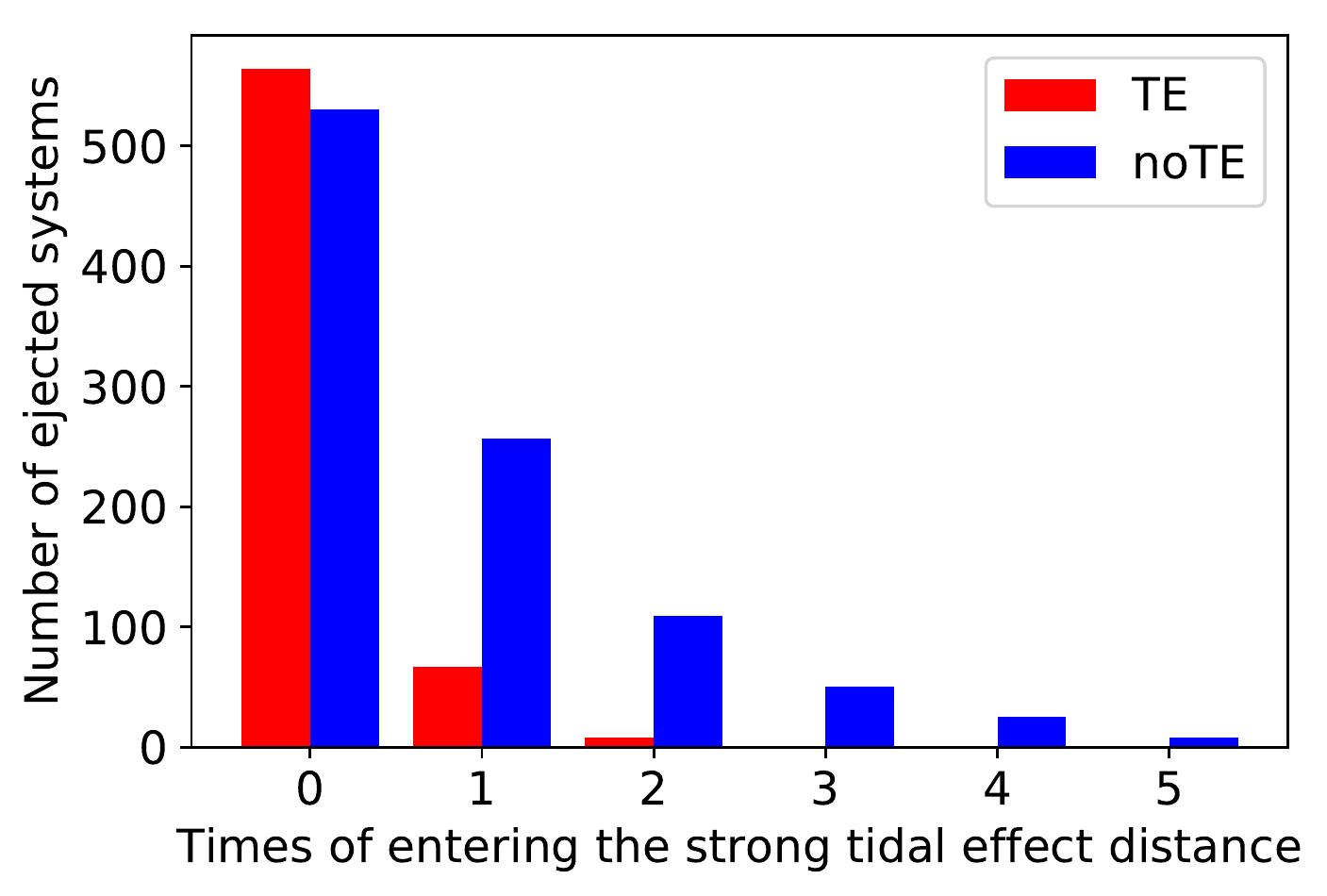}
    \caption{Categorization of the systems destabilized by ejections in terms of how many times the planets enter the strong tidal effect distance ($\rp < 4\RJ$) before the final ejection. The red and blue bars represent the data from the fiducial set of \texttt{TE} and \texttt{noTE} runs, respectively.}
    \label{fig:nbody-fid-nce}
\end{figure}

It is not surprising that the percentage of ejections decreases when the fluid effects are included. As the tides drain kinetic energy from the orbit, it is harder for the planets to reach the escape speed from the star. Fig.~\ref{fig:nbody-fid-nce} shows the number of encounters with $\rp<4\RJ$ in the systems that end with ejections. In the \texttt{noTE} runs, about half of the ejected planets undergo at least one such encounter. In the \texttt{TE} runs, however, planets that enter the $\rp<4\RJ$ escape channel are diverted to collisions by the tidal effects. Given the \texttt{noTE} data, by removing this portion of runs from the ejections and adding them to the mergers count, we can obtain (to a good accuracy) the branching ratio of the \texttt{TE} runs.

\subsubsection{Property of merger products}
\label{sec:nbody-fid-m}

\begin{figure}
    \includegraphics[width=\columnwidth]{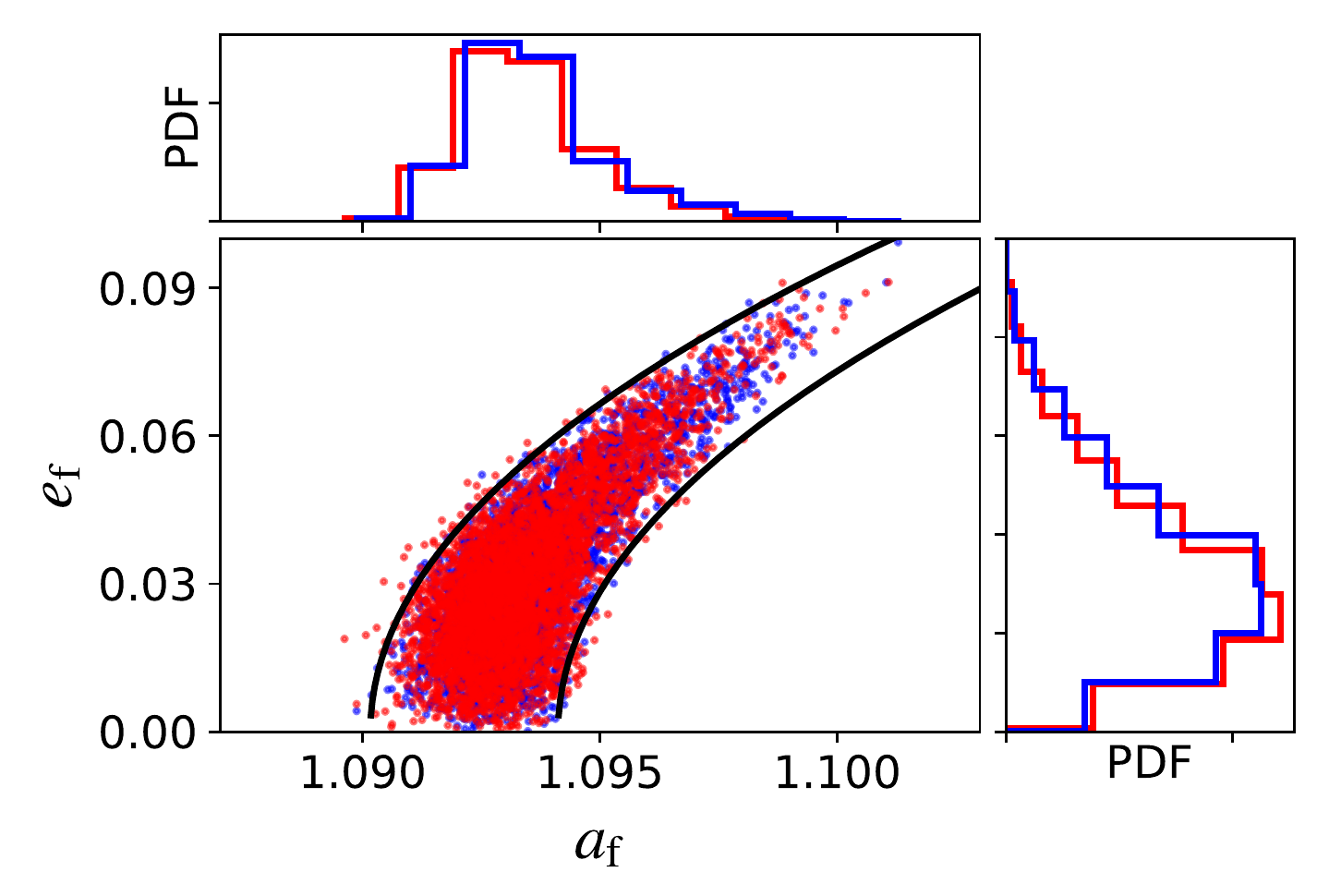}
    \caption{Eccentricity ($e_{\rm f}$) and semi-major axis ($a_{\rm f}$) of the merger products in the fiducial runs. The central panel shows the joint distribution of $e_{\rm f}$ - $a_{\rm f}$. The red dots are the merger products from the \texttt{TE} runs and the blue are from the \texttt{noTE} runs. The black lines are the $e_{\rm f}$-$a_{\rm f}$ relation (equation~\ref{eq:n-body-fid-coll-L}) due to the maximum and minimum possible angular momentum of the initial conditions. The histograms are the marginal distribution of $a_{\rm f}$ (top) and $e_{\rm f}$ (right).}
    \label{fig:nbody-fid-orb-m}
\end{figure}

Mergers between planets are treated as completely inelastic collisions in both the \texttt{noTE} and \texttt{TE} runs. Fig.~\ref{fig:nbody-fid-orb-m} shows the distribution of the semi-major axis ($a_{\rm f}$) and the eccentricity ($e_{\rm f}$) of the merger products in our fiducial simulations. Since the energy in the center-of-mass frame of the two planets is much smaller than their orbital energy around the star, the semi-major axis of a merger product can be estimated as
\begin{equation}
    \label{eq:n-body-fid-coll-af}
    a_{\rm f} \simeq \frac{(m_1+m_2)a_1a_2}{m_1a_2 + m_2a_1},
\end{equation}
using energy conservation. The estimated value of $a_{\rm f}$ for the fiducial runs is $1.08$AU, while the actual $a_{\rm f}$ from the simulations ranges between $1.09$ and $1.10$AU. Similar features were observed in \cite{Ford2001}. This ``discrepency'' is due to the extra energy released from the gravitational binding energy between the planets. The energy change due to the fluid effect during each close encounter is at least one order of magnitude smaller than the binding energy (Table~\ref{tab:fluid-fit}), so the fluid effect is not manifested in the final energy of the system.

The eccentricity of a merger product can be calculated from the conservation of angular momentum, which is mainly in the $\hat{z}$ direction, as
\begin{equation}
    \label{eq:n-body-fid-coll-L}
    \begin{split}
    L_{\rm z} = 
    & \mu_1\sqrt{M_{\star} a_1 (1-e_1^2)}\cos{I_1}+ \mu_2\sqrt{M_{\star} a_2 (1-e_2^2)}\cos{I_2} \\
    \simeq & \mu_{\rm f}\sqrt{M_{\star} a_{\rm f} (1-e_{\rm f}^2)},
    \end{split}
\end{equation}
where $\mu_i = m_iM_{\star}/(m_i+M_{\star})$. The maximum and minimum values of $e_{\rm f}$ (obtained using $e_1=e_2=0.05$ and $0.01$, respectively) as a function of $a_{\rm f}$ are plotted as the black lines in Fig.~\ref{fig:nbody-fid-orb-m}. These boundaries encloses the \texttt{noTE} results perfectly, but leave a small amount of \texttt{TE} data outside. The marginalized probability density functions of $e_{\rm f}$ and $a_{\rm f}$ shows that \texttt{TE} orbits are only slightly smaller and less eccentric.

\begin{figure}
    \includegraphics[width=\columnwidth]{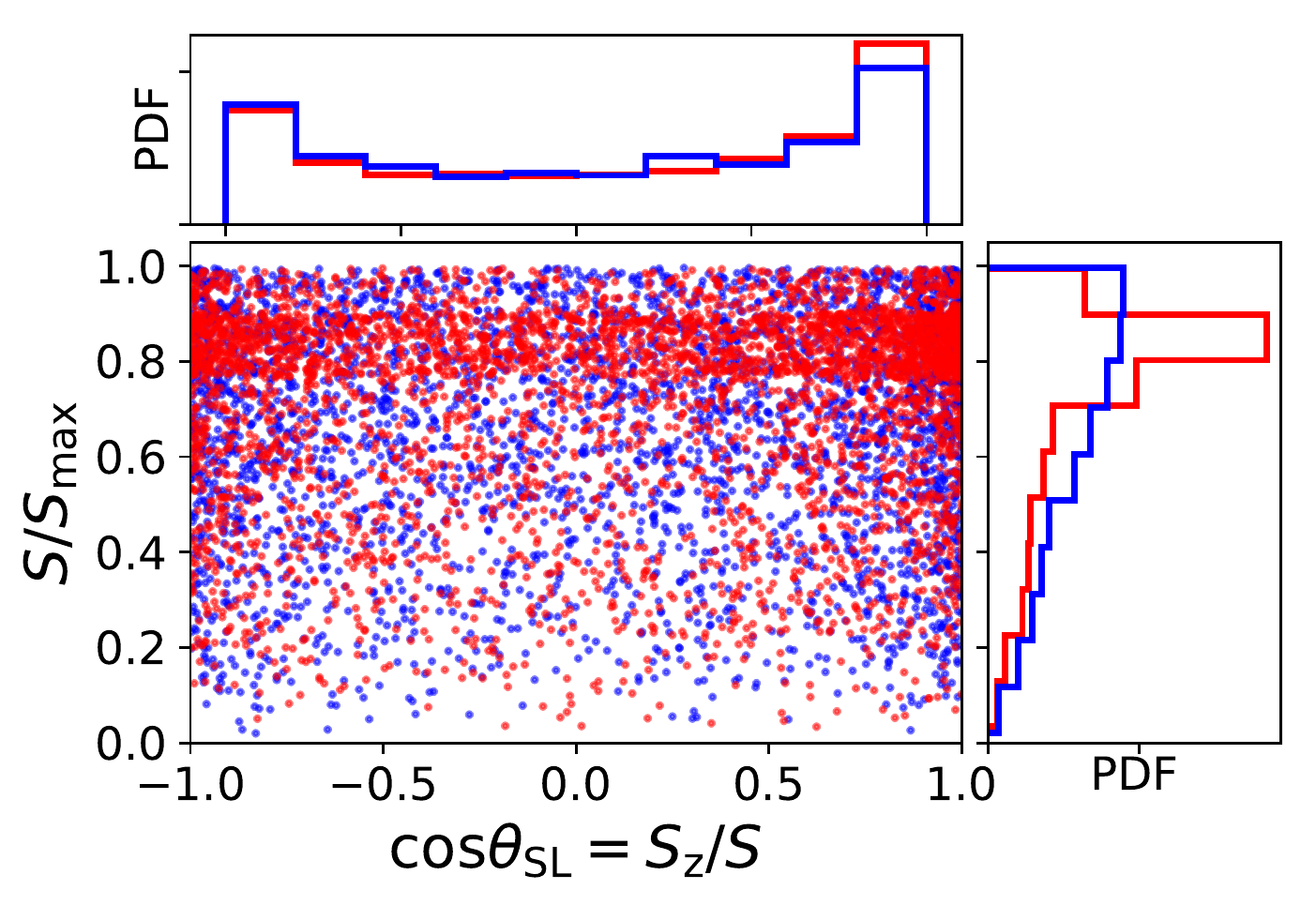}
    \caption{The spin magnitude and obliquity of the merger products found in our fiducial runs. The obliquity is displayed as $\cos\theta_{\rm SL}=S_z/S$ on the horizontal axis, and the spin is given in the unit of the maximum spin $S_{\rm max}=\mu\sqrt{2G(m_1+m_2)(R_1+R_2)}$ (where $\mu$ is the reduced mass of the two planets). The central panel shows the distribution in the $S/S_{\rm max}$ - $\cos\theta_{\rm SL}$ space. The red dots are from the \texttt{TE} runs and the blue are from the \texttt{noTE} runs. The histograms are the marginal distributions of $\cos\theta_{\rm SL}$ (top) and $S/S_{\rm max}$ (right).}
    \label{fig:nbody-fid-orb-S}
\end{figure}

In the center-of-mass frame of the planet ``binary'', the relative orbital angular momentum at the moment of collision turns into the spin of the merger product. The vertical axis of Fig.~\ref{fig:nbody-fid-orb-S} shows the resulting spin angular momentum, assuming no loss during the mergers and that the initial (pre-merger) spin  of each planet is negligible -- these assumptions are justified from our hydrodynamics simulations (Section~\ref{sec:fluid}) and the fact that the spins of the solar-system gas giants and those constrained for a few extrasolar planetary-mass objects are much smaller than the breakup value~\citep{Bryan2017}. This procedure also allows us to calculate the obliquity $\theta_{\rm SL}$ (the angle between the spin and orbital angular momentum axes) of the merger product (shown as the horizontal axis of Fig.~\ref{fig:nbody-fid-orb-S}). In \cite{Li2020a}, we carried out a detailed analysis of the spin and obliquity distributions from the \texttt{noTE} runs, including analytical predictions. In \texttt{TE} runs (the red dots and lines in Fig.~\ref{fig:nbody-fid-orb-S}), the spin magnitude distribution has a peak at $S\sim0.8 S_{\rm max}$, where 
\begin{equation}
\label{eq:Smax}
  S_{\rm max}=\mu \sqrt{2G(m_1+m_2)(R_1+R_2)},
\end{equation}
with $\mu=m_1m_2/(m_1+m_2)$ as the reduced mass of the two planets. In contrast, the \texttt{noTE} runs yield a spin distribution of $2S/S_{\rm max}$ \citep[see][]{Li2020a}. This discrepancy between \texttt{TE} and \texttt{noTE} is expected, as the systems that are ``tidally captured'' (i.e. those with the first-time encounter pericenter distance between $2\RJ$ and $4\RJ$) suffer tidal dissipation, and can merge in the following encounters with a smaller impact velocity. Fig.~\ref{fig:nbody-fid-orb-S} shows that  the obliquity distribution of the merger products is not affected by the fluid effects, and is almost the same for the \texttt{TE} and \texttt{noTE} runs. An approximated distribution for $\cos\theta_{\rm SL}$ is
\begin{equation}
    f_{\cos\theta_{\rm SL}} = \frac{1}{\pi}\frac{1}{\sqrt{1-\cos^2\theta_{\rm SL}}}.
\end{equation}
See \cite{Li2020a} for discussion of the regime of validity of this analytic distribution.

In summary, for our choice of initial conditions, giant planet mergers produce massive planets orbiting at $a_{\rm f} \simeq a_1$ and $e_{\rm f} \simeq [0.00, 0.09]$. Including the fluid effects shrink and circularize the orbit of a merger product by a very small amount. The merger products have a wide range of spin magnitudes and obliquities.

\subsubsection{Property of the ejection survivors}
\label{sec:nbody-fid-e}

\begin{figure}
    \includegraphics[width=\columnwidth]{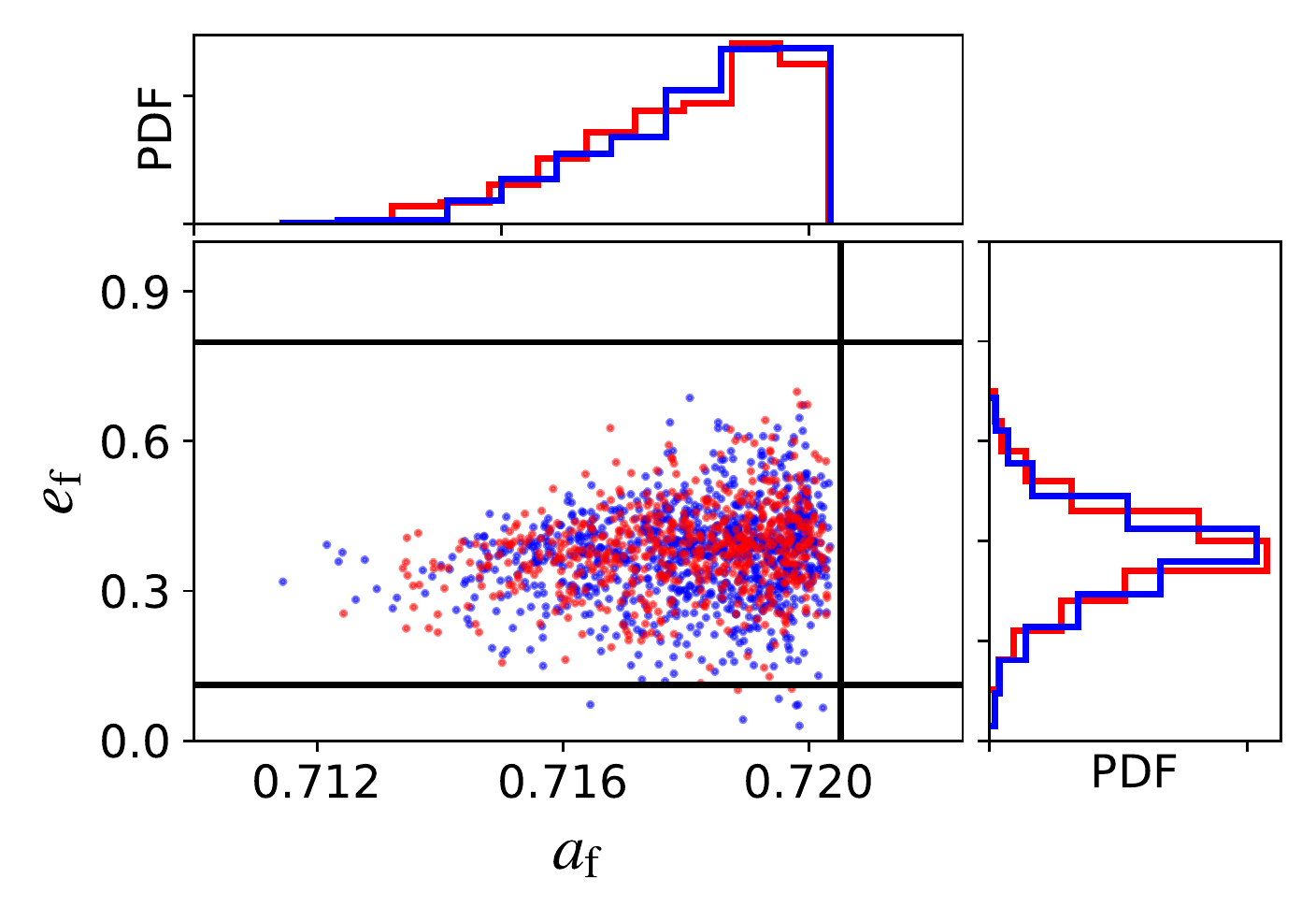}
    \caption{Eccentricity ($e_{\rm f}$) and semi-major axis ($a_{\rm f}$) of the remaining planets from the ejection events in the fiducial simulations. The central panel shows the joint distribution of $e_{\rm f}$ - $a_{\rm f}$. The red dots are the results from the \texttt{TE} runs and the blue are from the \texttt{noTE} runs. The black vertical line is the upper limit of $a_{\rm f}$ (equation~\ref{eq:nbody-fid-ejec-a}), and the two horizontal lines are limits of $e_{\rm f}$ obtained from equations~(\ref{eq:nbody-fid-ejec-L})-(\ref{eq:nbody-fid-ejec-e}). The histograms are the marginal distributions of $a_{\rm f}$ (top) and $e_{\rm f}$ (right).}
    \label{fig:nbody-fid-orb-e}
\end{figure}

With $m_1/m_2 = 2$, almost every ejection in our simulations has the lower mass planet being the runner. Since the leaving planet ($m_2$) carries small positive orbital energy to escape from the star, we know that the remaining planet ($m_1$) must have
\begin{equation}
    \label{eq:nbody-fid-ejec-a}
    a_{\rm f} \le \frac{m_1 a_1 a_2}{m_1a_2 + m_2a_1}
\end{equation}
from energy conservation. For the initial condition in our fiducial runs, this implies $a_{\rm f} \le 0.72$~AU. Let $a_{\rm esc}$, $e_{\rm esc}$ be the semi-major axis and eccentricity of $m_2$ before escaping but after the final close encounter. Angular momentum conservation requires
\begin{equation}
    \label{eq:nbody-fid-ejec-L}
    m_1\sqrt{a_{\rm f}(1-e_{\rm f}^2)} + m_2\sqrt{a_{\rm esc}(1-e_{\rm esc}^2)} \simeq
    m_1\sqrt{a_1} + m_2\sqrt{a_2}.
\end{equation}
On the other hand, the orbit crossing condition gives
\begin{equation}
    \label{eq:nbody-fid-ejec-e}
    a_{\rm f}(1+e_{\rm f}) \gtrsim a_{\rm esc}(1-e_{\rm esc}).
\end{equation}
Combining equations~(\ref{eq:nbody-fid-ejec-L})-(\ref{eq:nbody-fid-ejec-e}) and $1-e_{\rm esc}\ll 1$, we can solve for the allowed range of $e_{\rm f}$ as a function of $a_f$. This range is shown in Fig.~\ref{fig:nbody-fid-orb-e}.

Fig.~\ref{fig:nbody-fid-orb-e} shows the property of the remaining planets from the ejection events in our simulations. We see no significant difference between the results from \texttt{noTE} and \texttt{TE} runs. In Section~\ref{sec:nbody-fid-m}, we showed that most of the systems that reach $r_p\le 4R_J$ (and thus require the use of our fitting formulae for close encounters) end up as merger events. Hence, only a small fraction of the \texttt{TE} data for ejections are affected by the tidal effects.

\subsection{Results for different initial planet semi-major axes}
\label{sec:a1}

In the above (Section~\ref{sec:nbody-fid}), we have presented the results from our fiducial runs. Here we study how the results depend on the initial semi-major axes of the two planets. We adopt the same initial conditions as in the fiducial runs, but with the initial $a_1$ changing from $1$~AU to $10$~AU with $1$~AU increment for each set of runs.

\begin{figure}
    \includegraphics[width=\columnwidth]{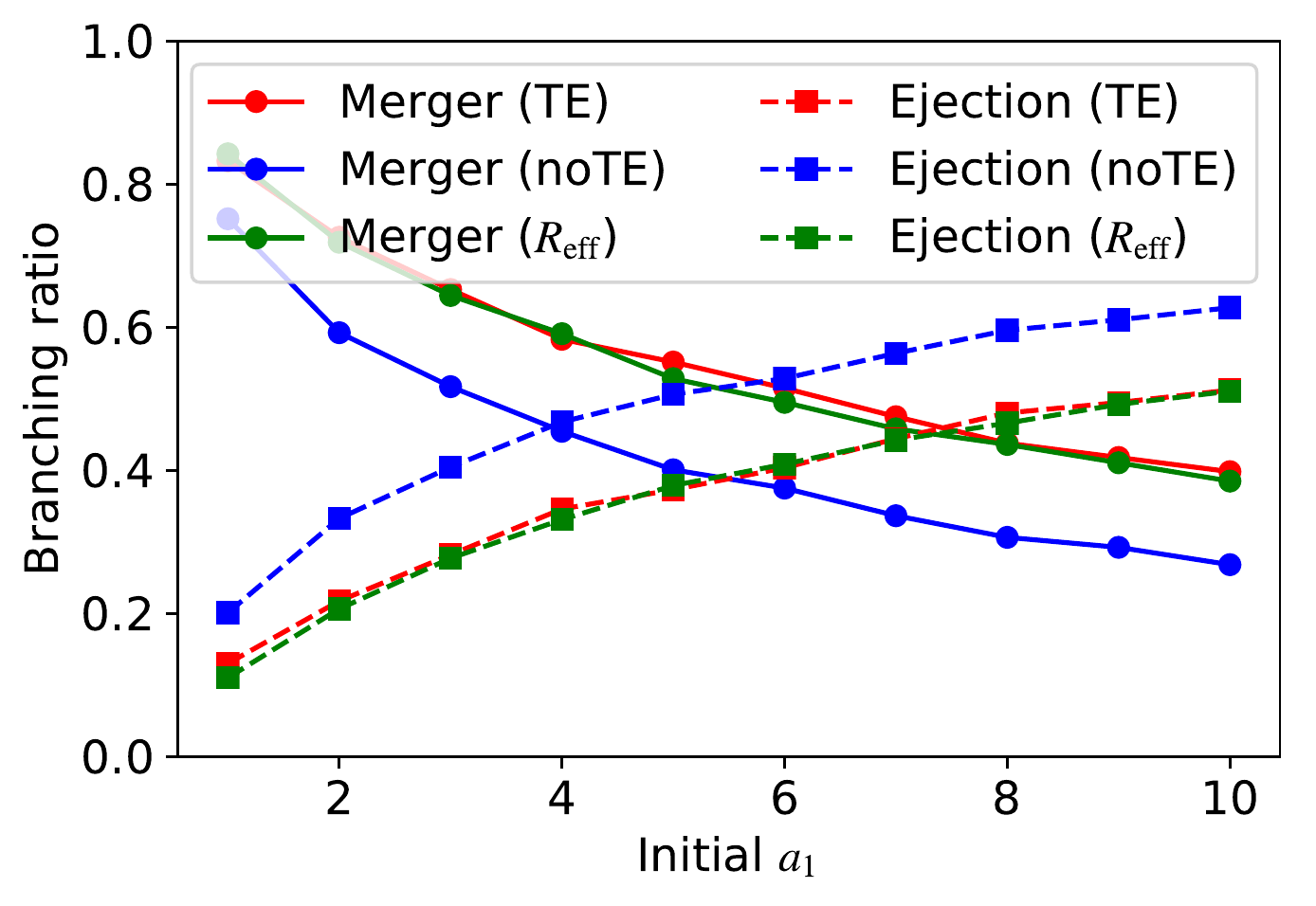}
    \caption{The branching ratio at $t=10^6 P_1$ as a function of initial $a_1$. The solid and dashed lines represent the fractions of systems destabilized by mergers and ejections, respectively. The red and blue are the results from the \texttt{TE} and \texttt{noTE} runs, while the green is calculated from the \texttt{noTE} results with equation~(\ref{eq:n-body-easy}).}
    \label{fig:nbody-asweep-br}
\end{figure}

In Fig.~\ref{fig:nbody-asweep-br}, we show the branching ratio as a function of the initial $a_1$. The decreasing merger fraction with increasing $a_1$ is consistent with the expection that planetary collisions are less likely as the Safronov number (the squared ratio of the escape velocity from the planetary surface to the planet’s orbital velocity) increases (e.g., \citealt{Ford2001,Petrovich2014,Anderson2020}) In essence, increasing the initial $a_1$ effectively reduces the radii of the planets, making the collisions less likely.

Similar to the fiducial runs, we find that the fluid effects on the orbital properties of the merger products and the ejection survivors are insignificant, and the results presented in Sections~\ref{sec:nbody-fid-m} and \ref{sec:nbody-fid-e} remain valid for general values of $a_1$.

\begin{figure}
    \includegraphics[width=\columnwidth]{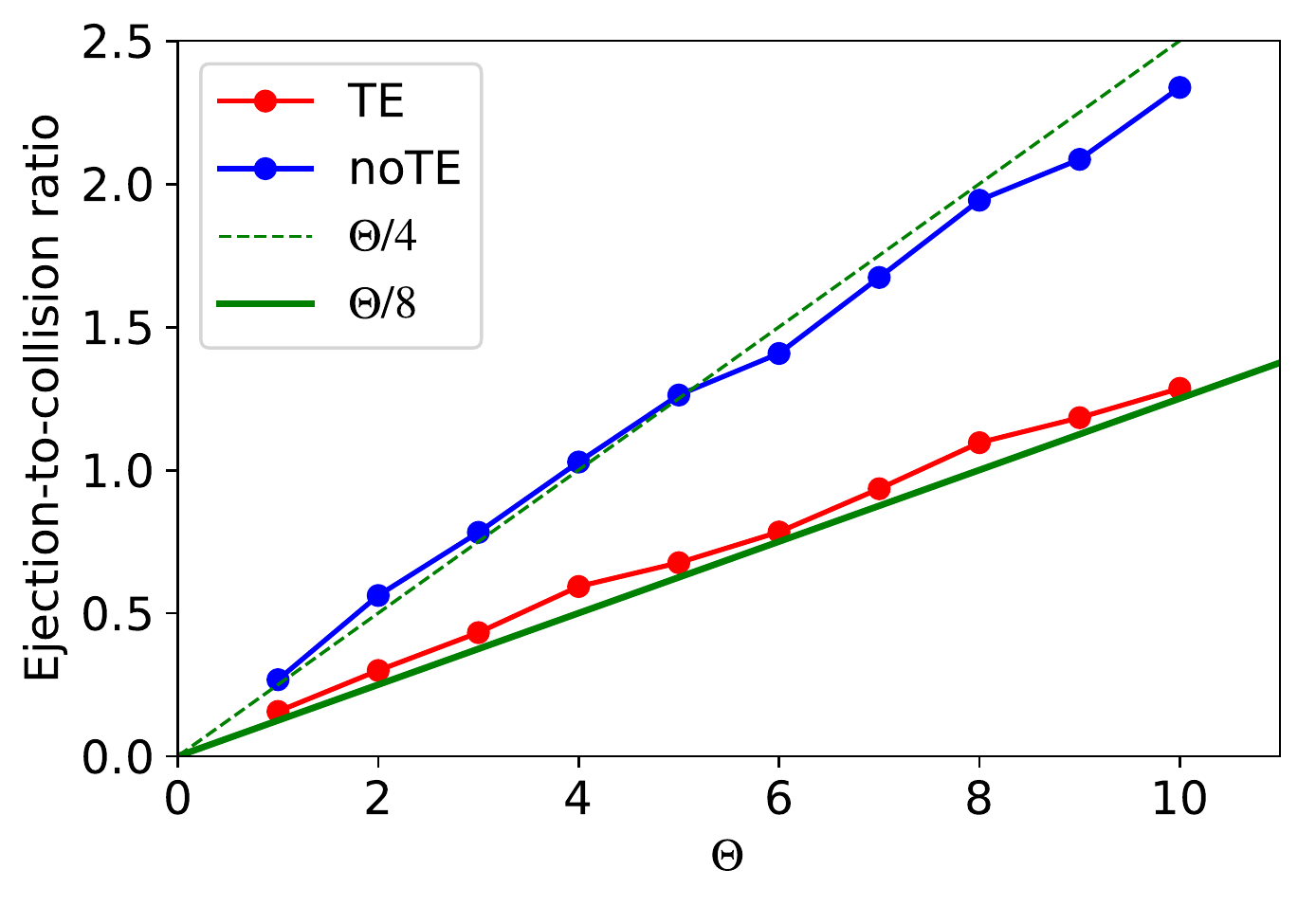}
    \caption{Ejection-to-collision ratio as a function of $\Theta \equiv ({R_{\rm p}}/{\RJ})^{-1}({a_1}/{1~\text{AU}})$ (which is proportional to the Safronov number). The data points from the \texttt{TE} and \texttt{noTE} runs fall on two straight (green) lines given by equation~(\ref{eq:saf-law}).}
    \label{fig:nbody-asweep-e2cr}
\end{figure}

Using the results from Section~\ref{sec:nbody-fid-br}, we may assume that the tidal effect will merge all planet binaries when their separation is less than $2R_{\rm eff}\equiv 4\RJ$. For the \texttt{noTE} runs, we monitor the number of systems that eject through the $\rp<4\RJ$ channel, denoted by $N_{4\RJ}$. By re-classifying all of them from ejections to mergers, the number of each outcome with the fluid effects can be estimated as
\begin{equation}
    \label{eq:n-body-easy}
    \begin{split}
    N_E(R_{\rm eff}) & \equiv N_E(\texttt{noTE}) - N_{4\RJ}, \\
    N_M(R_{\rm eff}) & \equiv N_M(\texttt{noTE}) + N_{4\RJ}, 
    \end{split}
\end{equation}
where $N_E$ and $N_M$ are the numbers of ejections and collisions, respectively. The branching ratios calculated using $N_E(R_{\rm eff})$ and $N_M(R_{\rm eff})$ are also plotted in Fig.~\ref{fig:nbody-asweep-br}. Not surprsingly, we find $N_E(R_{\rm eff})\simeq N_E(\texttt{TE})$ and $N_M(R_{\rm eff})\simeq N_M(\texttt{TE})$.

To quantify the dependence of the branching ratio on the planetary radius and the semi-major axis, we define the dimensionless ratio (proportional to the Safranov number)
\begin{equation}
  \Theta \equiv \left(\frac{R_{\rm p}}{\RJ}\right)^{-1} \left(\frac{a_1}{1~\text{AU}}\right).
\end{equation}  
Fig.~\ref{fig:nbody-asweep-e2cr} shows our numerical results for the ejection-to-collision ratio as a function of $\Theta$. We see that the ratio can be nicely fit by
\begin{equation}
    \label{eq:saf-law}
    \frac{N_E(\texttt{TE})}{N_M(\texttt{TE})} \simeq \frac{1}{2} \frac{N_E(\texttt{noTE})}{N_M(\texttt{noTE})} \simeq \frac{\Theta}{8},
\end{equation}
where this linear trend is a natural result of the competition between the gravitational focusing (assists collisions) and the random orbital energy drift (assists ejections) during close encounters \citep[see][]{Pu2020}. In this regard, we can consider planets in the \texttt{TE} runs as having an effective radius $R_{\rm eff}=2\RJ$ instead of $R_{\rm p}=\RJ$.

The findings described above suggest a simple prescription to account for the fluid effects in planet collisions: 
\begin{enumerate}
    \item When the separation between the two planets are less than $4\RJ$, merge the planets as a perfect inelastic collision.
    \item Otherwise, treat the planets as point masses.
\end{enumerate}
When dealing with a large ensemble of planetary system simulations, this prescription provides good estimates to both the branching ratio and the final orbital property of the planets.

\subsection{Compactness of the system}
\label{sec:K}

Here we examine how our results depend on the compactness of the two-planet systems. We use the same parameters as in the fiducial runs, but with the $K$ value (see equation~\ref{eq:KRH}) varying from $1.4$ to $2.5$ (all less than $2\sqrt{3}$, the critical value of Hill stability; see \citealt{Gladman1993}). Larger values of $K$ would require longer integration times to reach instability, so we do not consider $K > 2.5$ in this work.

\begin{figure}
    \includegraphics[width=\columnwidth]{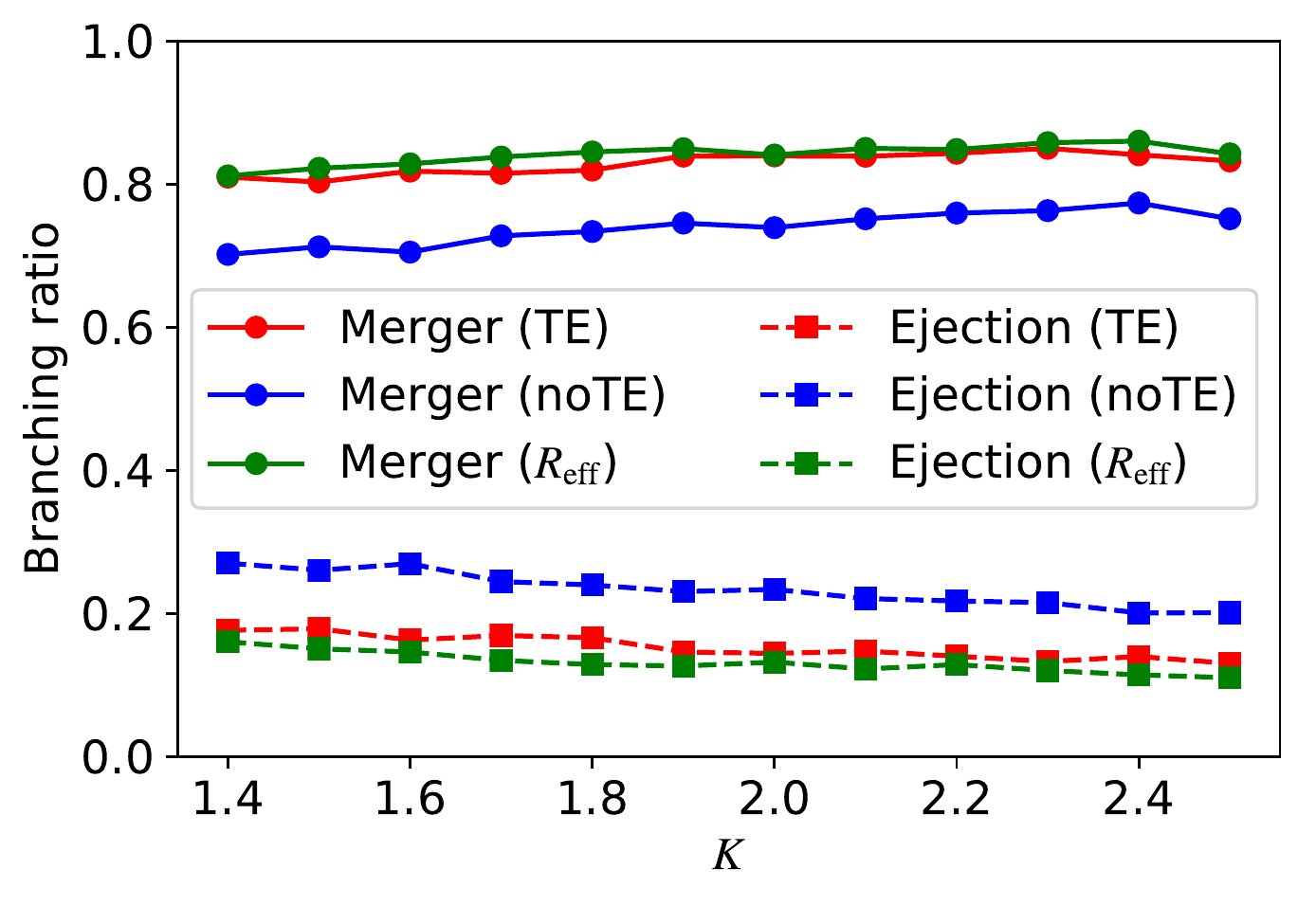}
    \caption{The branching ratio at $t=10^6 P_1$ as a function of the compactness parameter $K$ of the initial two-planet system. The solid and dashed lines represent the fractions of systems destabilized by collisions and ejections, respectively. The red and blue are the simulation results from the \texttt{TE} and \texttt{noTE} runs, while the green is calculated from the \texttt{noTE} results with equation~(\ref{eq:n-body-easy}).}
    \label{fig:nbody-ksweep-br}
\end{figure}

Fig.~\ref{fig:nbody-ksweep-br} shows that the branching ratios depend weakly on $K$, with more compact systems (small $K$'s) more likely to experience ejections. Equation~(\ref{eq:n-body-easy}) can be used to accurately predict the \texttt{TE} branching ratios from the \texttt{noTE} results for all $K$'s. The final distributions of $a$, $e$ and spin are similar to our fiducial results described in Section~\ref{sec:nbody-fid} (Figs~\ref{fig:nbody-fid-orb-m}-\ref{fig:nbody-fid-orb-e}).

\section{Conclusion}
\label{sec:conclusion}

We have studied the dynamical evolution of two giant planets, initially in nearly circular coplanar orbits, to determine the outcomes of close encounters/scatterings due to orbital instability. Although there already exists an extensive literature on this subject (see Section~\ref{sec:intro}), several issues related to this ``basic'' dynamics problem of two-planet scatterings have not been addressed adequately. Our paper extends previous works in several fronts:
\begin{enumerate}

\item  In the first part of this paper (Section~\ref{sec:fluid}), we perform hydrodynamics simulations (using SPH) of close encounters and collisions of two comparable-mass giant planets (each with radius $\Rp$ and modeled as a polytrope) to investigate the properties of the merger products and the bypassing planets in parabolic approaching orbits. We find that
\begin{enumerate}
    \item A one-shot merger of the planets happens when the impact parameter (the ``pericenter'' separation of point-mass planets), $\rp$, is less than the physical radius of the planet $\Rp$. A collision with $\rp$ between $\Rp$ and $2\Rp$ leads to an immediate loop-back of the binary planets and a merger during the second encounter.
    \item The merger products tend to be fast-spinning and puffy. They contain more than $97\%$ of the total mass from the initial planets. This also implies the conservation of momentum and angular momentum in mergers. Thus, giant planet mergers can be well described by perfect inelastic collisions.
    \item For larger impact parameters ($2\Rp<\rp<4\Rp$), the binary planets bypass each other with some mass exchange, orbital energy loss, change in eccentricity and apsidal advance happening near the pericenter. These effects diminish when $\rp$ is greater than $4\RJ$.  Combining with long-term orbital integrations (see below), we also find that, at least statistically, planet encounters with $\rp<4\Rp$ eventually lead to mergers.
\end{enumerate}

\item Based on our hydrodynamics simulations, we provide simple prescriptions (with fitting formulae) to take account of the fluid effects of close encounters between planets in $N$-body orbital simulations (Section~\ref{sec:prescription}).

\item We carry out a suite of two-giant-planet scattering simulations to determine the properties of various outcomes (Section~\ref{sec:nbody}). We find that 
\begin{enumerate}
    \item The fluid (tidal) effects significantly increase the branching ratio of planetary mergers relative to ejections. For typical giant planets ($M_1=2M_2=2\MJ$, $R_1=R_2=\RJ$), the merger fraction reaches $83\%$ for initial systems at $a_1=1$~AU and $40\%$ at $a_1=10$~AU (see Fig.~\ref{fig:nbody-asweep-br}). The branching ratio (with the fluid effects included) can be approximated by running standard ``sticky-sphere'' $N$-body simulations with an effective collision radius of $4\RJ$ (rather than $2\RJ$).  Our parameter study shows that this result is robust against varying initial $a_1$ and and the compactness parameter $K$ (defined in equation~\ref{eq:KRH}). The ejection-to-merger ratio can be well described by $(1/4)(R_{\rm eff}/\RJ)^{-1}(a_1/{\rm AU})$ (see Fig.~\ref{fig:nbody-asweep-e2cr}), and the fluid effects increase the effective radius $R_{\rm eff}$ from $\RJ$ to $2\RJ$.
    \item The fluid effects do not change the distributions of semi-major axis and eccentricity of each type of remnant planets (mergers vs surviving planets in ejections; see Figs~\ref{fig:nbody-fid-orb-m} and~\ref{fig:nbody-fid-orb-e}). However, since the branching ratios of mergers and ejections are changed, the overall distribution of orbital properties of planet scattering remnants are strongly affected by the fluid effects.
    \item The merger products have broad distributions of spin magnitudes and obliquities (Fig.~\ref{fig:nbody-fid-orb-S}). While the obliquity distribution is unchanged by the fluid effects \citep[see][]{Li2020a}, the distribution of $S$ exhibits a peak at $0.8S_{\rm max}$ due to the fluid effects (as opposed to $S_{\rm max}$ without the fluid effects; see equation~\ref{eq:Smax} for the definition of $S_{\rm max}$).
\end{enumerate}

To thoroughly explain the observed exoplanetary statistics, such as the eccentricity distribution, a much wider range of planetary system configurations needs to be considered. Although this work focus on planet pairs with a fixed mass ratio (mostly 2-to-1, and to a less extent 1.5-to 1 in Section~\ref{sec:fluid-mass-ratio}), it can affect the interpretations of the results in other studies that include more complex $N$-body systems \citep[see][]{Anderson2020}. Our result implies that, because of the larger planet merger fraction, it is more difficult to excite eccentricities via planet-planet scatterings, compared to the findings of previous works. On the other hand, Jupiter-like planets with larger masses and spins may be more common than in other $N$-body models. 

\end{enumerate}

\section*{Acknowledgements}
DL thanks the Dept. of Astronomy and the Miller Institute for Basic Science at UC Berkeley for hospitality while part of this work was carried out. KRA is supported by a Lyman Spitzer, Jr. Postdoctoral Fellowship at Princeton University. PB is supported by the National Aeronautics and Space Administration (NASA) on the NASA Earth and Space Sciences Fellowship. This work has been supported in part by the NSF grant AST-17152 and NASA grant 80NSSC19K0444. This paper makes use of the software packages \textsc{matplotlib} \citep{Hunter2007}, \textsc{numpy} \citep{Walt2011}, \textsc{REBOUND} \citep{Rein2012}, \textsc{SPLASH} \citep{Price2007}, and \textsc{StarSmasher} \citep{Gaburov2018}.

\section*{DATA AVAILABILITY}
The simulation data underlying this article will be shared on reasonable request to the corresponding author.




\bibliographystyle{mnras}




\appendix


\bsp	
\label{lastpage}
\end{document}